# Evidence of competing ground states between fractional Chern insulator and antiferromagnetism in moiré $MoTe_2$

Xumin Chang[1†], Feng Liu[1†], Fan Xu[1,2†], Cheng Xu[3,4†], Jiayong Xiao[1], Zheng Sun[1], Pengfei Jiao[1], Yixin Zhang[5], Shaozheng Wang[1], Bohan Shen[6], Renjie He[6], Kenji Watanabe[7], Takashi Taniguchi[8], Ruidan Zhong,[1,2] Jinfeng Jia[1,2,9,10], Zhiwen Shi[1,2], Xiaoxue Liu,[1,2,9] Yang Zhang[3,11*], Dong Qian[1,2*], Tingxin Li[1,2,9*] and Shengwei Jiang[1,2*]

[1]Key Laboratory of Artificial Structures and Quantum Control (Ministry of Education), School of Physics and Astronomy, Shanghai Jiao Tong University, Shanghai 200240, China

[2]Tsung-Dao Lee Institute, Shanghai Jiao Tong University, Shanghai, 201210, China

[3]Department of Physics and Astronomy, University of Tennessee, Knoxville, TN 37996, USA

[4]Department of Physics, Tsinghua University, Beijing 100084, China

[5]Max Planck Institute for Chemical Physics of Solids, 01187, Dresden, Germany

[6]Zhiyuan College, Shanghai Jiaotong University, Shanghai 200240, China

[7]Research Center for Electronic and Optical Materials, National Institute for Materials Science, 1-1 Namiki, Tsukuba 305-0044, Japan

[8]Research Center for Materials Nanoarchitectonics, National Institute for Materials Science, 1-1 Namiki, Tsukuba 305-0044, Japan

[9]Hefei National Laboratory, Hefei 230088, China

[10]Shanghai Research Center for Quantum Sciences, Shanghai 201315, China

[11]Min H. Kao Department of Electrical Engineering and Computer Science, University of Tennessee, Knoxville, Tennessee 37996, USA

[†]These authors contribute equally to this work.

[*]Emails: yangzhang@utk.edu, dqian@sjtu.edu.cn, txli89@sjtu.edu.cn, swjiang@sjtu.edu.cn

**Two-dimensional moiré materials present unprecedented opportunities to explore quantum phases of matter arising from the interplay of band topology and strong correlations[1–3]. One of the most striking examples is the recent observation of fractional quantum anomalous Hall (FQAH) effect in twisted bilayer $MoTe_2$ ($tMoTe_2$) with relatively large twist angles ($\theta \sim 3.7° - 3.9°$)[4–8]. The electronic ground states can be sensitive to the twist angle, as the twist angle determines the electron bandwidth and correlation strength in the moiré system. Here, we report the observation of unexpected competing ground states in $tMoTe_2$ moiré superlattice, on which balance can be tipped by both twist angle and electric field ($E$). Specifically, we observed evidence for anomalous antiferromagnetic (AFM) ground states with zero Hall resistance at both $\nu_h = 1$ and $2/3$, at intermediate twist angles $\theta \sim 3°$. The AFM orders are suppressed by applying vertical $E$. Emergent ferromagnetism accompanied by integer Chern insulator (ICI) or fractional Chern insulator (FCI) states are observed near the critical $E$ ($E_c$) of moiré superlattice symmetry transition, where the charge-transfer**

**mechanism is proposed as the mechanism of this ferromagnetism. Our results demonstrate *t*$MoTe_2$ as a fascinating platform for exploring unexpected correlated phases with nontrivial topology and fractional excitations and point to electric-field-controlled ultralow-power spin-valleytronic devices.**

## Main

The interplay between strong correlations and topology can lead to the emergence of intriguing quantum states of matter. Moiré materials based on transition metal dichalcogenides (TMDs) have emerged as a promising platform for exploring novel quantum phenomena that arise from band topology and many-body interactions[1–3,9–17]. Recently, there has been a surge in research focused on rhombohedral-stacked (R-stacked) *t*$MoTe_2$, in which transport experiments have provided unambiguous evidence of both QAH and FQAH effects on devices with relatively large twist angles $\theta \sim 3.7° - 3.9°$ [7,8]. Based on the obtained Chern numbers, photoluminescence and compressibility measurements suggest the FQAH is the ground state of $\nu_h = 2/3$ over a relatively large range of twist angle, specifically $\theta \sim 2.6° - 3.9°$, while QAH at $\nu_h = 1$ extends even down to $2.1°$[5,6,8,18–20]. The Chern numbers were obtained by comparing the Streda formula with the fitted dispersion slope of the filling factor versus the magnetic field ($B$) of ~Tesla order. However, the energy difference between the competing phases is usually subtle, and the applied relatively large $B$ could hide the true ground state at zero $B$. Transport measurement can provide unambiguous evidence of the true ground state as demonstrated in devices with relatively large twist angles ($\theta \sim 3.7° - 3.9°$), while transport study of $\nu_h \leq 1$ states in *t*$MoTe_2$ moiré device with small $\theta$ is challenging due to the increasing difficulty of making transparent electrical contact. On the other hand, in the moiré system, the ground states are usually expected to be sensitive to the twist angle[22], suggesting more unexpected correlated effects to explore in *t*$MoTe_2$ moiré system with smaller $\theta$[20–25].

In this work, we performed a systematic twist angle dependence study on the R-stacked *t*$MoTe_2$ moiré system by combining optical and transport measurements. We observed competing magnetic ground states depending on both twist angle and $E$. Specifically, we observed unexpected AFM ground states with vanishing Hall resistance at both $\nu_h = 1$ and $2/3$, at intermediate twist angles $\theta \sim 3°$, in between the FM/CI ground states of both larger and smaller $\theta$. Moreover, applying vertical electric field ($E$) induces an AFM to FM (ICI/FCI) transition and an emergent FM state was observed near critical electric field ($E_c$) of moiré superlattice symmetry transition.

## Twist-angle dependent magnetic ground states

Fig. 1a shows the schematic of our dual-gated device with a contact gate design for both optical and transport measurements, fabricated following the similar procedure reported earlier[8] (see the Methods). We have fabricated *t*$MoTe_2$ bilayer devices with twist angle $\theta \sim 2.4° - 4.1°$, denoted as $D(\theta)$. The application of $E$ can break energy degeneracy between the moiré orbitals localized in the MX (A) and XM (B) sublattices and induce layer polarization. $E$ dependence of RC measurement demonstrates a continuous tuning of hole distribution (Fig. 1c), from which $E_c$ of honeycomb-triangular superlattice symmetry transition can be determined (Extended Data Fig. 2).

To investigate twist angle dependence of magnetic interactions, we performed reflective magnetic circular dichroism (MCD) measurements under small $B$ on devices with representative twist angles (see the Methods for details of MCD measurements). Fig. 1d shows the MCD intensity plot as a function of $\nu_h$ and $E$ of D(4.1°). The magnetic response maximizes at $E = 0$ for $\nu_h \leq 1$. Increasing $E$ suppresses the MCD signal until it finally vanishes at large $E$. The MCD signal versus $B$ at $\nu_h = 1, 2/3$, and $E = 0$ are also measured (Fig. 1g). Sharp spin-flip transition and pronounced hysteresis loop, a hallmark of ferromagnetism, are observed. These results are consistent with the earlier report on devices with large twist angles ($\theta \sim 3.4° - 3.9°$)[4,5]. Fig. 1f shows the MCD map of a small $\theta$ device D(2.4°). Although the magnetic region shows a different pattern, the $E$ dependence of magnetic response shows a qualitatively similar trend as in D(4.1°), i.e., maximum at $E = 0$ and decreases with increasing $E$. The MCD signal versus $B$ at $\nu_h = 1$ and $E = 0$ are also measured (Fig. 1i). Appreciable hysteresis loop is observed, confirming its FM order.

Interestingly, the magnetic response shows substantially different behaviors in devices with intermediate $\theta \sim 3°$. Firstly, two hollow features around $E = 0$ at $\nu_h = 1$ and $2/3$ can be immediately identified in the MCD map of D(3.15°) (Fig. 1e). The strongest MCD response appears not at zero but finite $E$, signifying non-monotonic $E$ dependence of $\chi_{\mathrm{MCD}} = \frac{\partial MCD}{\partial B}$ (see the Methods for definition and detailed of MCD analysis). Secondly, continues spin-switching without appreciable hysteresis loop, i.e., no remnant magnetization at $B = 0$, is observed in the MCD-$B$ curves at $\nu_h = 1, 2/3$, at $E = 0$ (Fig. 1h). Thirdly, clear suppression of MCD response and a zero-MCD plateau is observed at small $B$ is observed at $\nu_h = 1$ and $E = 0$ (Fig. 1h). The above observations are in sharp contrast with large and small $\theta$ devices, suggesting the different magnetic ground states in *t*$MoTe_2$ devices with intermediate $\theta \sim 3°$ (see Extended Data Fig. 3 for additional data from D(3.0°) and D(3.1°)).

## Antiferromagnetism at $\nu_h = 1$ and $2/3$

To reveal the magnetic ground states of $\nu_h = 1$ and $2/3$ in devices with intermediate $\theta$, we performed systematic temperature ($T$) dependent MCD measurements. Fig. 2a shows the MCD map at various temperatures of D(3.15°). Remarkably, at $\nu_h = 1$ and $2/3$, while MCD minimums are observed at $E = 0$ at 1.6 K, their amplitudes gradually increase and evolve into maximums at elevated temperatures. That is to say, $\chi_{\mathrm{MCD}}$ of $\nu_h = 1$ and $2/3$ show non-monotonic dependence not only versus $E$, but also versus $T$ (see Extended Data Fig. 3 for additional data from D(3.0°) and D(3.1°)).

To see the $T$ dependence of $\chi_{\mathrm{MCD}}$ clearer, MCD versus $B$ at selective $T$ were measured at $\nu_h = 1, 2/3$ and $E = 0$ (Fig. 2b-c). The zero-field MCD slopes at $\nu_h = 1$ and $2/3$ first increase and then decrease with increasing $T$, showing clear non-monotonic $T$ dependence of $\chi_{\mathrm{MCD}}$. Similar behavior is also observed in D(3.0°) (Extended Data Fig. 4). Moreover, we observed clear suppression of $\chi_{\mathrm{MCD}}$ at small $B$ at low $T$, at $\nu_{\mathrm{h}} = 1$ (Fig. 2b), which gradually disappeared with elevated $T$. The non-monotonic $E$ and $T$ dependences here are in sharp contrast to devices with larger ($\theta \geq 3.4°$) or smaller ($\theta \leq 2.7°$) twist angles, in which monotonic $E$ and $T$ dependences are observed both in our measurements (Extended Data Fig. 3) and earlier reports[4–6]. The non-monotonic $\chi_{\mathrm{MCD}} - T$ dependence and the suppression of $\chi$ at small magnetic are characteristic features of antiferromagnetism[26], suggesting AFM ground states of D(3.15°) at $\nu_{\mathrm{h}} = 1$ and $2/3$ at $E = 0$.

Fig. 2d shows $R_{\mathrm{xy}}$ versus $B$ at selective $T$ of $\nu_{\mathrm{h}} = 1$ at $E = 0$. We observed no Hall signal at zero $B$, and clear suppression of Hall slope at 1.6 K. The Hall suppression gradually disappeared with elevated $T$, showing non-monotonic $T$ dependence of zero-field $R_{\mathrm{xy}}$ slope, consistent with the MCD observation. Moreover, $R_{\mathrm{xy}}$ saturates to the quantized value of $\frac{h}{e^2}$ at critical $B$ ($B_{\mathrm{c}}$), demonstrating a magnetic field-induced AFM-ICI phase transition. It is worth noting that $T = 7$ K shows the best quantization at finite $B$, i.e., $B_{\mathrm{c}}$ is the smallest, compared with both lower and higher $T$, consistent with the expectation of the AFM ground state of $\nu_{\mathrm{h}} = 1$ at $E = 0$. Fig. 2e shows $R_{\mathrm{xy}}$ versus $B$ at selective $T$ of $\nu_{\mathrm{h}} = 2/3$ at $E = 0$. The observations are qualitatively similar to $\nu_{\mathrm{h}} = 1$, while $R_{\mathrm{xy}}$ saturates to the quantized value of $\frac{3h}{2e^2}$ at finite $B_{\mathrm{c}}$, demonstrating a magnetic field-induced AFM-FCI phase transition. Again, we find $B_{\mathrm{c}}$ is the smallest at temperature $\sim 2.5 - 3.5$ K, consistent with the expectation of the AFM ground state of $\nu_{\mathrm{h}} = 2/3$ at $E = 0$.

## Emergent ferromagnetism and Chern states

Fig. 3a and 3b show the MCD versus $B$ as a function of $E$ at $\nu_{\mathrm{h}} = 1$ and $2/3$, respectively. The zero-field slopes at $\nu_{\mathrm{h}} = 1$ and $2/3$ are initially enhanced by increasing $E$, reaching maximums at $\sim E_{\mathrm{c}}$, and then suppressed by further increasing $E$, showing clear non-monotonic $E$ field dependences (see Extended Data Fig. 4 for additional data from D(3.0°)). Surprisingly, we observed appreciable hysteresis loop emerged at ~ $E_{\mathrm{FM}}$ slight smaller than $E_{\mathrm{c}}$ of superlattice symmetry transition, signifying the FM order. Extended Data Fig. 5a and 3c show the MCD versus $B$ at selective $T$ at $\nu_{\mathrm{h}} = 1, 2/3$ at $E_{\mathrm{FM}}$, respectively. The small-field slopes decrease with increasing $T$, showing monotonic $T$ dependence of $\chi_{\mathrm{MCD}}$. The monotonic $\chi_{\mathrm{MCD}} - T$ dependences at $E \sim E_{\mathrm{c}}$ are in sharp contrast to the non-monotonic $\chi_{\mathrm{MCD}} - T$ dependences observed at $E = 0$, consistent with the FM order of $\nu_{\mathrm{h}} = 1$ and $2/3$ at $E_{\mathrm{FM}}$. The above result demonstrates a pure electric-field-induced AFM-FM transition, which has never been observed in moiré systems and any other material systems, to the best of our knowledge. The emergence of FM order near $E_{\mathrm{c}}$ also reflects the deep connection between the FM exchange mechanism and electronic lattice symmetry.

We also analyze the nature of the magnetic states at representative electric fields by an Arrott plot (that is, $M^2$ versus $H/M$, using MCD in place of $M$) in Extended Data Fig.

6. Here a curve with a positive low-field slope and positive y-intercept is FM-ordered, that with a positive x-intercept and a positive slope is PM, and that with a negative low-field slope is AF-ordered.[27–30] The Mott state (the $E = 29\ \mathrm{mV\,nm^{-1}}$ curve) is PM. Although the superexchange interaction, as reflected by a negative $T_{\mathrm{w}}$ (Extended Data Fig. 9), favors an AFM-ordered state, the geometric frustration from a triangular lattice greatly suppresses the ordering temperature. This behavior is in contrast to the $E = 0\ \mathrm{mV\,nm^{-1}}$ state. Here the negative low-field slope in the Arrott plot shows its AFM-ordering, which directly correlates with the suppressed $\chi_{\mathrm{MCD}}$ at small magnetic fields. Because geometric frustration is absent in the bipartite honeycomb lattice at small $E$, and finite temperature AF-ordering is favored. The slop sign change at large magnetic field reflects the field polarizing process. At intermediate $E$ (the $E = 23\ \mathrm{mV\,nm^{-1}}$ curve), a positive slope with positive y-intercept is observed, signifying the emergent FM-order.

Inspired by the above observation in the MCD measurements, we further performed transport measurements in device D(3.15°) to examine the topological properties of this system as a function of $E$. Fig. 3d show $R_{\mathrm{xy}}$ and $R_{\mathrm{xx}}$ versus $B$ at selective $E$ fields at $\nu_{\mathrm{h}} = 1$, $T = 1.5$ K. A Qualitatively similar trend is observed, consistent with the MCD measurements. No Hall signal is observed at zero $B$, while $R_{\mathrm{xy}}$ saturates to the quantized value of $\frac{h}{e^2}$ at $B_{\mathrm{c}}$ of $\sim 200$ mT at $E = 0$. $B_{\mathrm{c}}$ decreases with increasing $E$, down to $\sim 100$ mT at $E_{\mathrm{FM}}$ of $\sim 23\ \mathrm{mV\,nm^{-1}}$. An appreciable hysteresis loop is observed, confirming its FM order. Another interesting feature is $R_{\mathrm{xx}}$ shows an anomalous positive magnetoresistance (MR) at small $B$ at $E = 0$, and the positive MR gradually evolves into a negative MR at larger $E$ near $E_{\mathrm{c}}$[4,31].

Fig. 3e shows $R_{\mathrm{xy}}$ and $R_{\mathrm{xx}}$ versus $B$ at selective $E$ fields at $\nu_{\mathrm{h}} = 2/3$ of D(3.15°) at $T = 1.5$ K. We observed no Hall signal at zero $B$, while $R_{\mathrm{xy}}$ saturates to the quantized value of $\frac{3h}{2e^2}$ at $B_{\mathrm{s}} \sim 200$ mT at $E = 0$, and $B_{\mathrm{c}}$ decreases with increasing $E$, down to $\sim 20$ mT at around $E_{\mathrm{FM}}$ of $\sim 11\ \mathrm{mV\,nm^{-1}}$. A small but resolvable hysteresis loop is observed, while it is much more pronounced in MCD measurement. We attribute the discrepancy to the fact that the optical measurement is a more local measurement with sub-micro-sized light spot (See methods). In contrast, the transport measurement involves a much larger area (Extended Data Fig.1), thus more affected by spatial inhomogeneity. The spatial inhomogeneity effect combined with the smaller exchange energy scale of $\nu_{\mathrm{h}} = 2/3$ could smear out the intrinsic hysteresis effect.

$R_{\mathrm{xx}}$ also shows an anomalous positive MR at small $B$ at small $E$ at $\nu_{\mathrm{h}} = 2/3$. Similar to $\nu_{\mathrm{h}} = 1$, the positive MR gradually evolves into a negative MR at larger $E$. The observation of positive MR at small $B$ at $\nu_{\mathrm{h}} = 1$ and $2/3$ at $E = 0$ also excludes one possible artifact: the absence of MCD and Hall response here is caused by the formation of nano-sized FM domains, i.e., superparamagnetic state, which could smooth and smear out the magnetic hysteresis. If that is the case, we would expect to see negative MR near $B = 0$ (or coercive field if there is hysteresis), as applying $B$ polarizes the magnetic domains and helps the development of chiral edge states, as observed in all QAH and FQAH systems[7,8,17,31–33]. Therefore, we conclude that the anomalous positive

MR at small $B$ is closely correlated with the AFM order. The sign inversion of MR at large $B$ signifies a $B$-induced AFM-ICI(FCI) transition. On the other hand, we indeed observed negative MR at all temperatures at $E \sim E_c$ at $\nu_h = 2/3$ (Fig. 3f), which is consistent with the typical behavior of the QAH and FQAH, confirming the ICI and FCI are the zero-field ground states of $\nu_h = 1$ and $2/3$ at $\sim E_{FM}$, respectively.

The combined optical and transport measurements above show that the magnetic order of $\nu_h = 1$ and $2/3$ are closely connected to their topological properties. It's expected to show more robust FM order and possibly better zero-field quantization near $E_c$ at lower temperatures. However, the worsened device contact at lower temperatures prevents us from performing reliable magneto-transport measurements at mK temperature, which could be improved with better device quality in the future.

To study the electronic ground state of $E = 0$ state, we performed dielectric sensing measurements on this system, which effectively measures the compressibility of the system.[34]. Here we show the 2s sensing signal of $WSe_2$ versus $\nu_h$ at $E = 0$ in Extended Data Fig.7. A pronounced feature of resonance-energy-shift is observed at $\nu_h = 1$, signifying an incompressible state at $E = 0$. Unfortunately, we didn't see resolvable features of $\nu_h = 2/3$, likely due to its much smaller gap which is beyond the sensitivity of our measurement.[6] By the combined transport and dielectric sensing measurement, we conclude the $E = 0$ state at $\nu_h = 1$ is an insulator with a small gap (likely <6 K).

**Magnetic phase diagrams**

To map out the magnetic phase diagram of the system as a function of $T$ and $E$, we performed $T$ and $E$-dependent MCD measurements. Fig. 4a and 4b shows the MCD map and corresponding $\chi_{MCD} - T$ curves at representative $E$ at $\nu_h = 1$ of D(3.15°). Non-monotonic $\chi_{MCD} - T$ behavior with a clear $\chi_{MCD}$ peak at around 7 K is observed at $E = 0$. We assign the peak temperature to be Neel temperature ($T_N$), where the AFM order starts to develop[26]. Similar non-monotonic $\chi_{MCD} - T$ behaviors are also observed at finite $E$ fields below $E_c$, while $T_N$ decrease with increasing $E$. On the other hand, $\chi$ increases monotonically with decreasing $T$ at $E$ fields above $E_c$, suggesting a different magnetic nature. Qualitatively similar behavior is also observed at $\nu_h = 2/3$, except $T_N$ are significantly lower, $\sim 3.1\ K$ at $E = 0$, and $E_c$ of $\sim 11$ mV nm$^{-1}$, as shown in Fig. 4c and 4d (see Extended Data Fig. 8 for additional data from D(3.0°)). We also performed $T$ and $E$-dependent $R_{xy}$ measurements at $\nu_h = 1$ and $2/3$ in D(3.15°) at small $B$ (Extended Data Fig. 7f and 11a). Non-monotonic $R_{xy} - T$ dependencies with $R_{xy}$ peak temperature decreases with increasing $E$, are observed at $E$ fields below $E_c$. For $E > E_c$, the sample becomes too insulating, preventing reliable $R_{xy}$ measurements. The results are consistent with the MCD measurements, confirming the non-monotonic $\chi_{MCD} - T$ dependence at small $E$ is an intrinsic property of $t$MoTe$_2$ with intermediate $\theta$.

The magnetic phase diagrams are mapped out as a function of $E$ and $T$ based on the $\chi_{MCD} - T$ dependences. For $E < E_c$, $T_N$ is determined from the peak temperatures in $\chi_{\mathrm{MCD}} - T$ curves[26], denoted by the arrows. For $E > E_{\mathrm{c}}$, Curie–Weiss temperature $T_{\mathrm{w}}$ can be extracted by analyzing $\chi_{\mathrm{MCD}} - T$ curves using Curie–Weiss law[4,17,35,36]. At high temperatures, the inverse MCD slope follows the Curie–Weiss law, ${\chi_{\mathrm{MCD}}}^{-1} \propto T - T_{\mathrm{w}}$ (dash lines) reasonably well (See Extended Data Fig. 9 for Curie–Weiss analysis). $T_{\mathrm{w}}$ is proportional to the effective spin-spin exchange interaction[37], and a negative (positive) $T_{\mathrm{w}}$ reflects an AFM (FM) exchange interaction. The $E$ dependences of extracted $T_{\mathrm{N}}$ and $T_{\mathrm{w}}$ of $\nu_{\mathrm{h}} = 1$ and $2/3$ of D(3.15°) are summarized in Fig. 4e and 4f. $T_{\mathrm{N}}$ of ~7 K is obtained at $E = 0$, which decreases monotonically with increasing $|E|$. Negative $T_{\mathrm{w}}$ of $\sim -9$ K at $E = 0$ is obtained from Curie–Weiss fitting, in reasonable agreement with $T_{\mathrm{N}}$. This shows the system is AFM ordered at small $E$, and the AFM order can be suppressed by applying $E$. For $E > E_{\mathrm{c}}$, negative $T_{\mathrm{w}}$ are obtained, showing a paramagnetic phase with AFM interaction, consistent with the expectation of antiferromagnetic kinetic exchange in a half-filled triangular lattice, as the earlier report in large twist angle device[4]. A positive maximum $T_{\mathrm{w}} \sim 3.3$ K is obtained at $\nu_{\mathrm{h}} = 1$ around $E_{\mathrm{c}}$, coincides with the $\chi_{\mathrm{MCD}}$ peak at 1.6 K, and is consistent with the FM hysteresis loop observed around $E_{\mathrm{c}}$ (Fig. 3a). $\nu_{\mathrm{h}} = 2/3$ shows qualitatively similar behavior as $\nu_{\mathrm{h}} = 1$, while the exchange energy scale is significantly smaller. The maximum $T_{\mathrm{N}}$ is ~3 K at $E = 0$, while $T_{\mathrm{w}}$ maxima around $E_{\mathrm{c}}$ is ~2.7 K. The above results demonstrate a cascade of $E$-induced magnetic and topological phase transitions from AFM order ($E < E_{\mathrm{c}}$) to FM-ordered ICI/FCI ($E \sim E_{\mathrm{c}}$), and finally to paramagnetic phases ($E > E_{\mathrm{c}}$).

The pronounced zero Hall/MCD plateau and the non-monotonic $\chi - T$ behavior with a peak at finite $T$ are fully consistent with the AFM ground state at $E = 0$, while the inplane FM state can be ruled out: $\chi_{\mathrm{MCD}}$ of inplane FM should be constant all the way until saturation when the magnetic field is applied along the hard axis (our-of-plane direction), and its $\chi_{\mathrm{MCD}} - T$ behavior should just saturate at $T < T_{\mathrm{c}}$ (see section 7 of the supplementary materials for detailed discussion).

## Discussion and conclusion

At zero electric field, strong Chern ferromagnetism has been observed at both large angle $\theta \sim 3.4° - 4.1°$ and small angle $\theta < 2.1° - 2.4°$. However, the direct Chern ferromagnetism exchange appeared weak at the intermediate twist angle, leading to an AFM order at $\theta \sim 3°$. We notice the zero Hall plateau (at $\nu_{\mathrm{h}} = 1$), the positive MR at small $B$ and its temperature evolution, and the $B$-induced AFM to CI transition looks quite similar to the even-layer $MnBi_2Te_4$ system[38]. Theoretically, the axion insulator state could be realized if the topological surface states of a three-dimensional topological insulator are gapped out by FM order on the surface with magnetizations pointing inward or outward[39], as in even-layer $MnBi_2Te_4$[40–43]. However, in our $t$$MoTe_2$ system, the two $MoTe_2$ layers are believed to be strongly hybridized, thus, should be treated as a two-dimensional system. Further theoretical and experimental investigations are needed to reveal the full nature of the AFM phase, but it's beyond the scope of the current study.

At $\theta \sim 3.0°$, the topmost two moiré bands in twisted $MoTe_2$ can be described by a two-band effective tight-binding model, where the Wannier orbits form a honeycomb lattice in real space[44]. Due to the maximized layer polarization of the Wannier orbits, an applied $E$ induces an on-site energy difference between these orbits. RC measurement demonstrates a continuous $E$ tuning of hole distribution on the two sublattices (Fig. 1c). Starting from the large $E$ limit at $\nu_h = 1$, the charges are polarized on one sublattice, forming a triangular lattice. The spin-spin interaction on the triangular lattice is dominated by the effective antiferromagnetic kinetic exchange interaction $J \sim t_2^2/U$ from virtual hopping mediated by double occupancy. The geometric frustration prevents colinear AFM order, resulting in the ground state of 120° Néel order (triangular AFM) below $T_N$, which is much lower than $T_W$. The negative $T_W$ and absence of AFM order in the large $E$ limit at $\nu_h = 1$ could be understood by the above mechanism and is consistent with earlier reports on homo- and hetero-bilayer TMD moiré systems[4,16]. The negative $T_W$ and absence of AFM order at $\nu_h = 2/3$ is also observed in $WSe_2/WS_2$ hetero-bilayer moiré system[36], consistent with our observation at $\nu_h = 2/3$ and large $E$, which could also be understood by the kinetic exchange mechanism.

Upon doping the half-filled triangular lattice, in the large $U$ limit of the Hubbard model, the well-known Nagaoka ferromagnetism typically occurs[35]. However, we observed the FM order at $\nu_h = 1$ in our experiment. On the other hand, in our experimental parameter region, the calculation shows the system remains in an AFM state at $\nu_h > 1$ under large $E$ limit, rulling out Nagaoka ferromagnetism. The AFM-ordered ground state at $E = 0$ also excludes the Chern ferromagnetism/inter-site Hund's coupling as the dominating FM exchange mechanism here, since it should be strongest at $E = 0$. Therefore, a different exchange mechanism is needed here to understand the FM phase near $E_c$. In the bipartite Honeycomb lattice, we propose a charge transfer mediated ferromagnetic exchange mechanism similar to Nagaoka ferromagnetism to describe this phenomenon. At finite $E$, when the on-site energy difference $\Delta$ is slightly smaller than the Hubbard $U$ but much larger than the interlayer tunneling near $\nu_h = 1$, the hole charge excitations occupy the B sublattice to avoid high-energy double-occupancy states in the A sublattice. The charge transfer drives the majority of holes in A sublattice into ferromagnetic states, enhancing the hopping between the B and A sublattices, resulting in a greater kinetic energy gain and lowering the system's energy, as shown in Extended Data Fig.12. The above charge-transfer mechanism explains the emergence of FM order at finite $E$ slightly smaller than $E_c$ (see section 1 of the supplementary materials for detailed discussion). A more detailed theoretical analysis can be found in our separated theoretical paper[45]. We believe the emergence of FM order at finite $E$ at $\nu_h = 2/3$ could also be understood by a similar mechanism.

In summary, we observed evidence for anomalous AFM ground states in the $t$$MoTe_2$ at intermediate twist angles θ~3° at $\nu_h = 1$ and $2/3$, by the combined optical and transport measurements. We have also observed the competing ground states between fractional Chern insulator and AFM, on which balance can be tipped by both twist angle and electric field. Our work unveils the rich competing ground states and quantum phase transitions in the $t$$MoTe_2$ system and paves the way for exploring correlated states with nontrivial topology and fractional excitations.

## Methods

### Device fabrication

Devices were fabricated using the standard tear-and-stack and dry transfer method. In all devices, to facilitate good contact formation for hole doping in $t\mathrm{MoTe_2}$, we use few-layer $TaSe_2$ as contact electrodes, taking advantage of its high work function. Thin flakes of hexagonal boron nitride ($h$-BN), graphite (3– 5 nm), monolayer $MoTe_2$ and $WSe_2$(HQ Graphene), and $TaSe_2$ (2– 5 nm) were first exfoliated from bulk crystals onto Si/$SiO_2$ (285 nm) substrates. Flakes of appropriate thickness were identified according to their reflectance contrast under an optical microscope. The $MoTe_2$ monolayer was mechanically cut into two parts using AFM tips. We handled $MoTe_2$ flakes inside a nitrogen-filled glovebox with oxygen and water levels below 0.01 ppm to minimize the degradation of $MoTe_2$. The thickness of $h$-BN flakes is also measured using an atomic-force microscope (AFM). The complete dual-gated structure was assembled using a polycarbonate stamp within the $N_2$ glovebox environment. A high-precision rotation stage (Thorlabs PRM1Z8) controls the targeted twist angle with a typical accuracy of ~0.1°. For optical devices, the stack was then released on the pre-patterned and cleaned gold electrodes of ~20 nm thickness, such that the graphite gates and $TaSe_2$ grounding contact were electrically connected to the deposited contacts. For dielectric sensing devices, the moiré bilayer and the $WSe_2$ monolayer are separated by an ultrathin $h$-BN layer with a thickness of approximately 1 nm. The complete dual-gated allows for full $h$-BN encapsulation of the air-sensitive $MoTe_2$ layers. The stamp polymer was dissolved in anhydrous chloroform for 5 minutes in ambient conditions (see Extended Data Fig.1). For transport devices, to achieve ohmic contacts irrespective of the applied electric fields, we fabricated the dual-gated devices with $SiO_2$/Si global contact gates to heavily hole-dope the contact regions, where $t\mathrm{MoTe_2}$ overlaps with few-layer $TaSe_2$ flakes. To define the Hall bar geometry of the $t\mathrm{MoTe_2}$ channel and eliminate the unavoidable monolayer $MoTe_2$ region, we carried out extra EBL and RIE processes, following a procedure similar to the one reported earlier[8]. The dual-gate geometry allows independent control of $\nu_\mathrm{h}$ and $E$ by the top gate voltage $V_\mathrm{t}$ and bottom gate voltage $V_\mathrm{b}$.

### Optical measurements

The optical measurements were performed in reflection geometry in a home-built confocal optical microscope system based on a closed-cycle helium cryostat (base temperature 1.6 K) equipped with a superconducting magnet (9 T). A superluminescent light-emitting diode with a peak wavelength of 1070 nm and full width at half-maximum bandwidth of 90 nm was used as the light source. The output of the diode was coupled to a single-mode fiber and focused onto the device under normal incidence by a low-temperature microscope objective (Attocube, 0.8 numerical aperture). The spot size of the excitation light, defined as the FWHM of the diffraction-limited beam spot, can be calculated as $\mathrm{FWHM} = 0.51\frac{\lambda}{\mathrm{NA}} \sim 0.7\ \mu\mathrm{m}$. A combination of a linear polarizer and an achromatic quarter-wave plate was used to generate the left and right circularly polarized light. To avoid perturbing the system, the incident intensity on the sample was kept below $30\ \mathrm{nW}\ \mu\mathrm{m}^{-2}$ (see section 3 of the supplementary materials for more information). The reflected light of a given helicity was spectrally resolved by a spectrometer coupled to a liquid nitrogen-cooled InGaAs one-dimensional array sensor (Princeton Instruments PyLoN-IR 1.7). The reflectance contrast spectrum was obtained by comparing the reflected light spectrum from the sample with the reference spectrum measured on a heavily doped condition (which is featureless in the spectral region of interest).

In Fig. 1c, the two crossed optical resonances on the right side of the neutral-exciton-like resonance disperse almost linearly with $E$-field, which is a typical behavior of Stark shift of interlayer exciton transitions, as also reported in earlier work[4]. The other two optical resonances on the left side of the layer-polarized exciton resonance are attributed to the bonding and anti-bonding branch of the layer-hybridized exciton resonances, as reported in earlier work[6].

The MCD spectrum is defined as $\mathrm{MCD}(E) = \frac{R^+(E)-R^-(E)}{R^+(E)+R^-(E)}$ , where $R^+(E)$ and $R^-(E)$ denote the reflection intensity spectrum of the left and right circularly polarized light. To analyze the MCD as a function of tuning parameters such as $\nu$, $E$, $B$, and $T$, we integrate the MCD modulus over a spectral range that covers most of the spectrum range of the MCD signal. We have performed integration that covers different ranges of the MCD spectrum, and the results are qualitatively the same, confirming the validity of our analysis (see Figure S5). The integrated MCD (referred to simply as the MCD below) reflects the difference in occupancy between the K and K′ valleys in $MoTe_2$. Because of spin-valley locking, the signal is proportional to the out-of-plane magnetization[4,16,35,36,46]. For simplicity, the MCD measured under a constant small $B$ ($\chi_{\mathrm{MCD}}$) is approximated as proportional to magnetic susceptibility $\chi = \frac{\partial M}{\partial B}$ (see section 4 of the supplementary materials for detailed discussion). In the MCD map and MCD susceptibility measurements, the results are averaged by positive and negative $B$, i.e., $\mathrm{MCD} = \frac{\mathrm{MCD}(+B)-\mathrm{MCD}(-B)}{2}$.

For dielectric sensing measurement, a halogen lamp serves as a white light source, the output of which is first collected by a single-mode fiber and collimated by a ×10 objective. The beam on the sample has a power below 1 nW.

## Determination of doping density and electric field

The carrier density $n$ and electric field $E$ on the sample are converted from the top (bottom) gate voltage $V_\mathrm{t}$ ($V_\mathrm{b}$) using a parallel plate capacitor model: $n = \frac{V_\mathrm{t}C_\mathrm{t}+V_\mathrm{b}C_\mathrm{b}}{e}$ and $E = \frac{1}{2}\left(\frac{V_\mathrm{t}}{d_\mathrm{t}} - \frac{V_\mathrm{b}}{d_\mathrm{b}}\right) - E_{\mathrm{offset}}$, where $e$ is the electron charge, and $\varepsilon_0$ is the vacuum permittivity and $C_\mathrm{t}$ and $C_\mathrm{b}$ are the top and bottom gate capacitance obtained from the top and bottom $h$-BN thickness $d_\mathrm{t}$ and $d_\mathrm{b}$ measured by atomic force microscopy (Asylum Research Cypher S). The offset electric field likely arises from the structure asymmetry and is determined from the symmetric axis of the dual-gate RMCD map.

The obtained doping density from the capacitor model can then be used to calculate the twist angle from the assigned filling factors in the reflection contrast measurements in optical devices.

For transport devices, the twist angle can be more accurately calibrated from the $R_{xx}$ map and Landau fan diagrams. The Extended Data Fig.10 presents the $B$ and filling dependences of $R_{xx}$. The Landau fan diagram, denoted by dashed lines, highlights a series of minima in $R_{xx}$ corresponding to the different Landau levels. Upon complete filling of $N$th Landau levels with hole carriers, the hole carrier density ($n_h$) can be expressed by $n_h = N|B|/\phi$, where $\phi = h/e$ ($h$ is Planck's constant). The hole carrier density difference ($\Delta n_h$) between neighboring Landau levels is given by $\Delta n_h = B/\phi$ . By determining the number of filling factors corresponding to these hole carrier densities, moiré density can be obtained. When $B$ reaches 10 Tesla, a change of 0.086 in the filling factor is observed, so the moiré density $n_M$ is established to be $2.8 \times 10^{12}$ cm$^{-2}$ . Finally, the twist angle $\theta$ is calculated to be 3.15° using $\theta = a\sqrt{\frac{\sqrt{3}}{2} n_M}$, where $a$ denotes the $MoTe_2$ lattice constant, approximately 3.5 Å.

**Transport measurements**

Electrical transport measurements were performed in a closed-cycle $^4$He cryostat (Oxford TeslatronPT) equipped with a 12 T superconducting magnet. Electrical transport measurements were performed using standard low-frequency lock-in techniques. The bias current is limited within 2 nA to avoid the disturbance of fragile quantum states. Finite longitudinal-transverse coupling occurs in our devices that mixes the longitudinal resistance $R_{xx}$ and Hall resistance $R_{xy}$. To correct this effect, we used the standard procedure to symmetrize $\left[\frac{R_{xx}(B)+R_{xx}(-B)}{2}\right]$ and anti-symmetrize $\left[\frac{R_{xy}(B)-R_{xy}(-B)}{2}\right]$ the measured $R_{xx}$ and $R_{xy}$ under positive and negative magnetic fields to obtain accurate values of $R_{xx}$ and $R_{xy}$, respectively. The measurements were carried out without any optical excitation.

**Acknowledgments**

We thank Fengcheng Wu, Xiaoyan Xu, and Mingpu Qin for fruitful discussions. Research was primarily supported by the National Key R&D Program of China (Nos. 2021YFA1401400, 2021YFA1400100, 2022YFA1405400, 2022YFA1402702, 2022YFA1402404, 2019YFA0308600, 2022YFA1402401, 2020YFA0309000), the National Natural Science Foundation of China (Nos. 12174250, 12141404, 12350403, 12174249, 92265102, 12374045), the Innovation Program for Quantum Science and Technology (Nos. 2021ZD0302600 and 2021ZD0302500), the Natural Science Foundation of Shanghai (No. 22ZR1430900). S.J., T.L. and X.L. acknowledge the Shanghai Jiao Tong University 2030 Initiative Program. T.L. and S.J. acknowledge the Yangyang Development Fund. Yang Zhang acknowledges support from AI-Tennessee and Max Planck partner lab grant. K.W. and T.T. acknowledge support from the JSPS KAKENHI (Nos. 21H05233 and 23H02052) and World Premier International Research Center Initiative (WPI), MEXT, Japan.

**Author contributions**

S.J., T.L., and D.Q. designed the scientific objectives and oversaw the project. F.L., J.X., and X.C fabricated the devices. X.C, J.X, and S.W. performed the optical measurements. F.X. and Z.S. performed the transport measurements. X.C., J.X., F.X., T.L., and S.J. analyzed the data. P.J. provided assistance in the optical measurements. C.X., Yixin Zhang, and Yang Zhang performed theoretical studies. R.Z. grew the bulk hBN crystals. K.W. and T.T. grew the bulk hBN crystals. S.J. and Yang Zhang wrote the manuscript. All authors discussed the results and commented on the manuscript.

## Figures

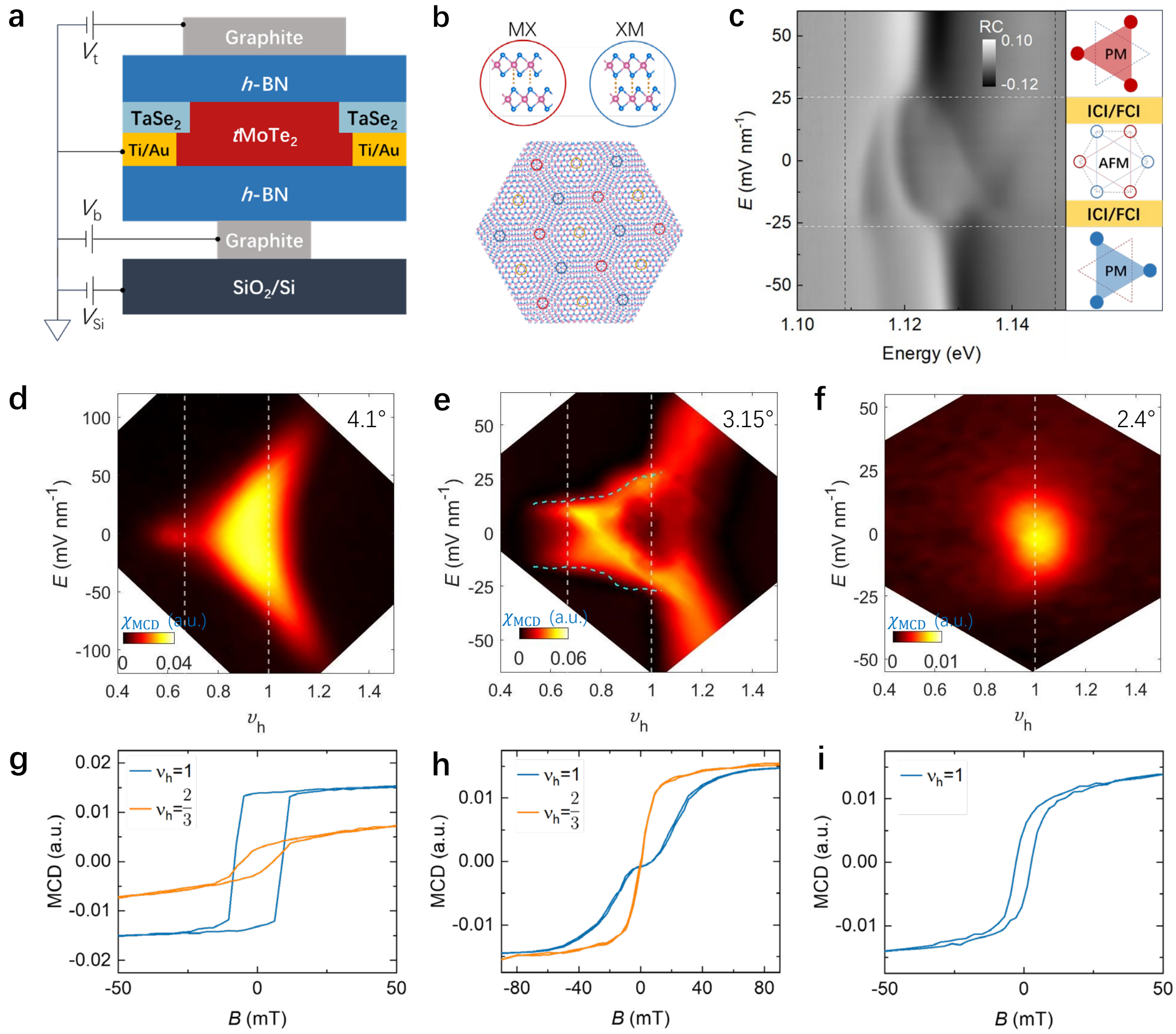

**Fig. 1 | Magnetic and electrostatic characterization of *t*MoTe₂.**

**a,** Schematic of a dual-gated device with contact gate design for optical and transport measurements. $V_t$, $V_b$ and $V_{si}$ are the bias voltages applied to the top graphite/hBN gate, the bottom graphite/hBN gate, and the $SiO_2$/Si global contact gate, respectively. **b,** The schematic moiré superlattices of R-stacked *t*$MoTe_2$. High-symmetry stackings are highlighted by circles. The red circles correspond to MX sublattice sites, where the Mo atoms in the top layer are aligned with the Te atoms in the bottom layer. Blue circles are the corresponding XM sublattice sites. MX and XM are two degenerate energy minima at $E = 0$, forming a honeycomb moiré superlattice. **c,** Optical reflectance contrast (RC) spectrum as a function of $E$ at $\nu_h = 1$ of D(3.15°). The charged exciton features disperse with $E$ in the low $E$ range, signifying the layer-hybridized region, which is an effective honeycomb lattice. At large $E$, the dispersive features disappear, and a new exciton feature appears which does not disperse with $E$, signifying the charges are fully polarized into one layer, forming a triangular lattice[4,6]. The white dashed lines denote $E_c$ that separate the layer-hybridized and layer-polarized regions. Chern insulator (CI) and fractional Chern insulator (FCI) are observed near $E_c$. **d-f,** MCD maps as a function of $\nu_h$ and $E$ of *t*$MoTe_2$ devices with twist angles of 4.1°, 3.15°, and 2.4° under a small $B$ (9 mT, 10 mT, 5 mT), respectively. $\nu_h = 1$ and $2/3$ are indicated by the vertical dashed lines. Measurements were done at 1.6 K. The cyan dashed lines denote $E_c$ of superlattice symmetry transition. **g-i,** The corresponding Magnetic-field dependence of the MCD at $\nu_h = 1$ and $2/3$. Spontaneous MCD and magnetic hysteresis are observed in D(4.1°) and D(2.4°) but not in D(3.15°).

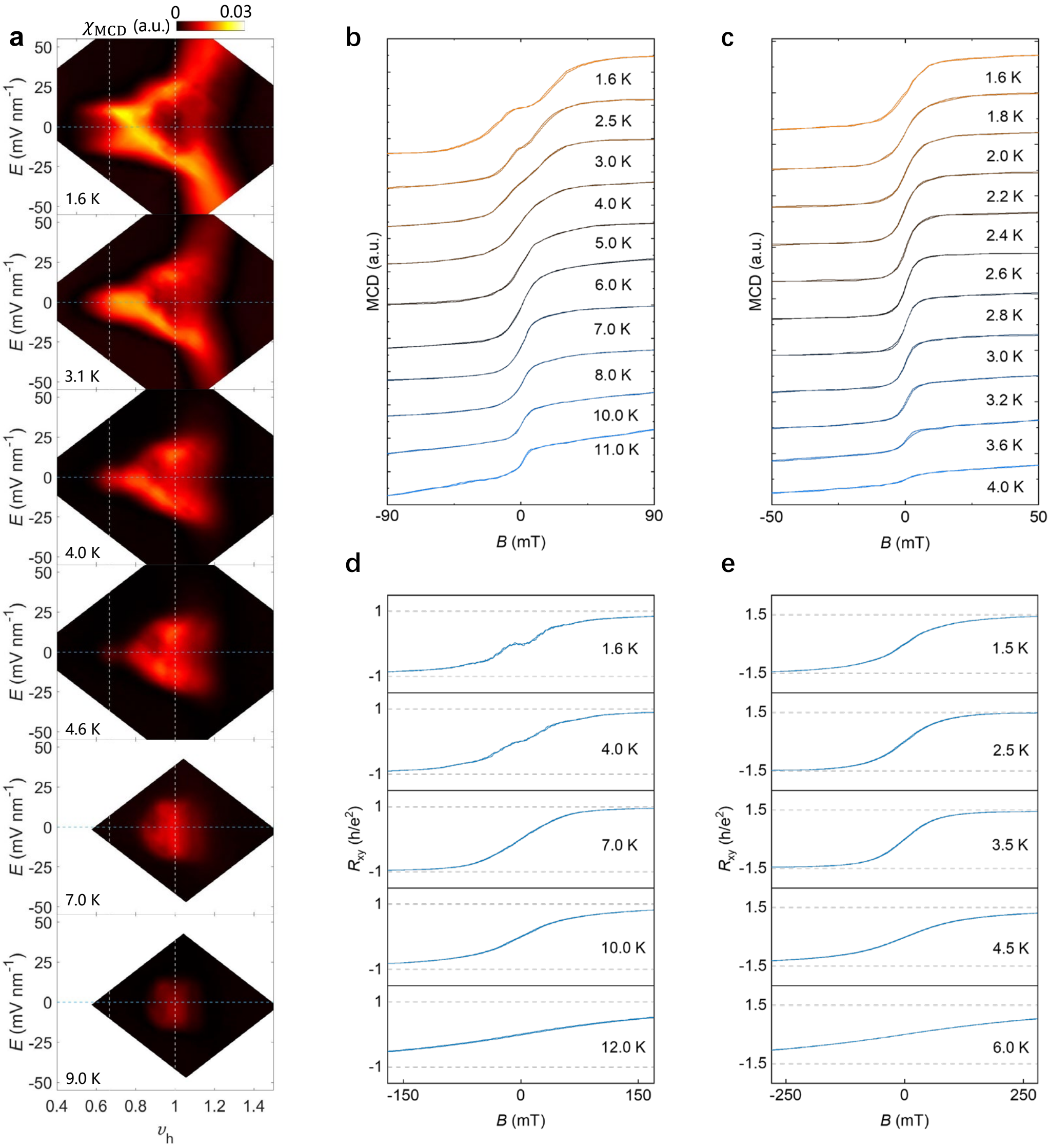

**Fig. 2 | Magnetic ground states at $\nu_h = 1$ and $2/3$.**

**a,** Evolution of MCD map as a function of $T$. Each color map plane shows the MCD versus $\nu_h$ and $E$ at a fixed $T$, under a small $B$ (~10 mT). $\nu_h = 1$ and $2/3$ are indicated by the white dashed lines. $E = 0$ is indicated by the blue dashed line. **b,c,** MCD versus $B$ at representative $T$ at $\nu_h = 1$ (**b**) and $2/3$ (**c**) at $E = 0$, respectively. The zero-field slopes of both fillings first increase and then decrease with increasing $T$, showing non-monotonic $T$ dependence of MCD $\chi$. **d,e,** $B$ dependence of $R_{xy}$ at representative temperatures at $\nu_h = 1$ (**d**) and $\nu_h = 2/3$ (**e**) at $E = 0$ mV nm$^{-1}$. The quantized values are indicated by the dashed lines.

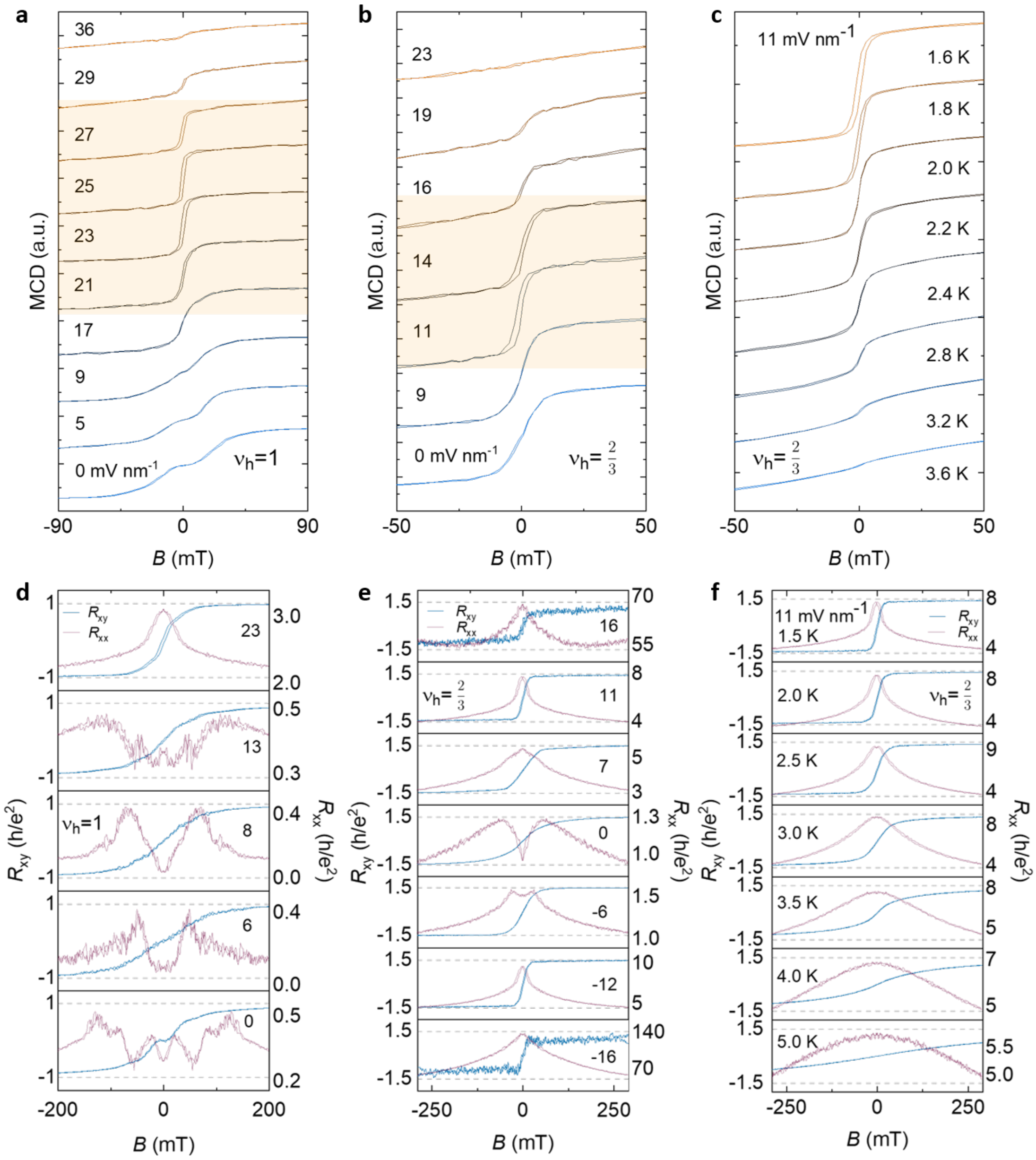

**Fig. 3 | Emergent CI and FCI phases near superlattice symmetry transition.**

**a,b,** $B$ dependence of MCD at representative electric fields (unit in $\mathrm{mV\,nm^{-1}}$) at $\nu_h = 1$ (**a**) and $2/3$ (**b**) of D(3.15°), respectively. The FM regions are highlighted by the yellow backgrounds. **c,** MCD versus $B$ at representative $T$ at $\nu_h = 2/3$ and $E = 11\ \mathrm{mV\,nm^{-1}}$. The zero-field slope decreases monotonically with increasing $T$. **d,e,** $B$ dependence of $R_{xy}$ and $R_{xx}$ at representative electric fields at $\nu_h = 1$ (**d**) and $2/3$ (**e**) of D(3.15°), respectively. The quantized values are indicated by the dashed lines. $R_{xy}$ and $R_{xx}$ are measured at $T = 1.5$ K. **f,** $B$ dependence of $R_{xy}$ and $R_{xx}$ at representative temperatures at $\nu_h = 2/3$ and $E = 11\ \mathrm{mV\,nm^{-1}}$. The quantized values are indicated by the dashed lines.

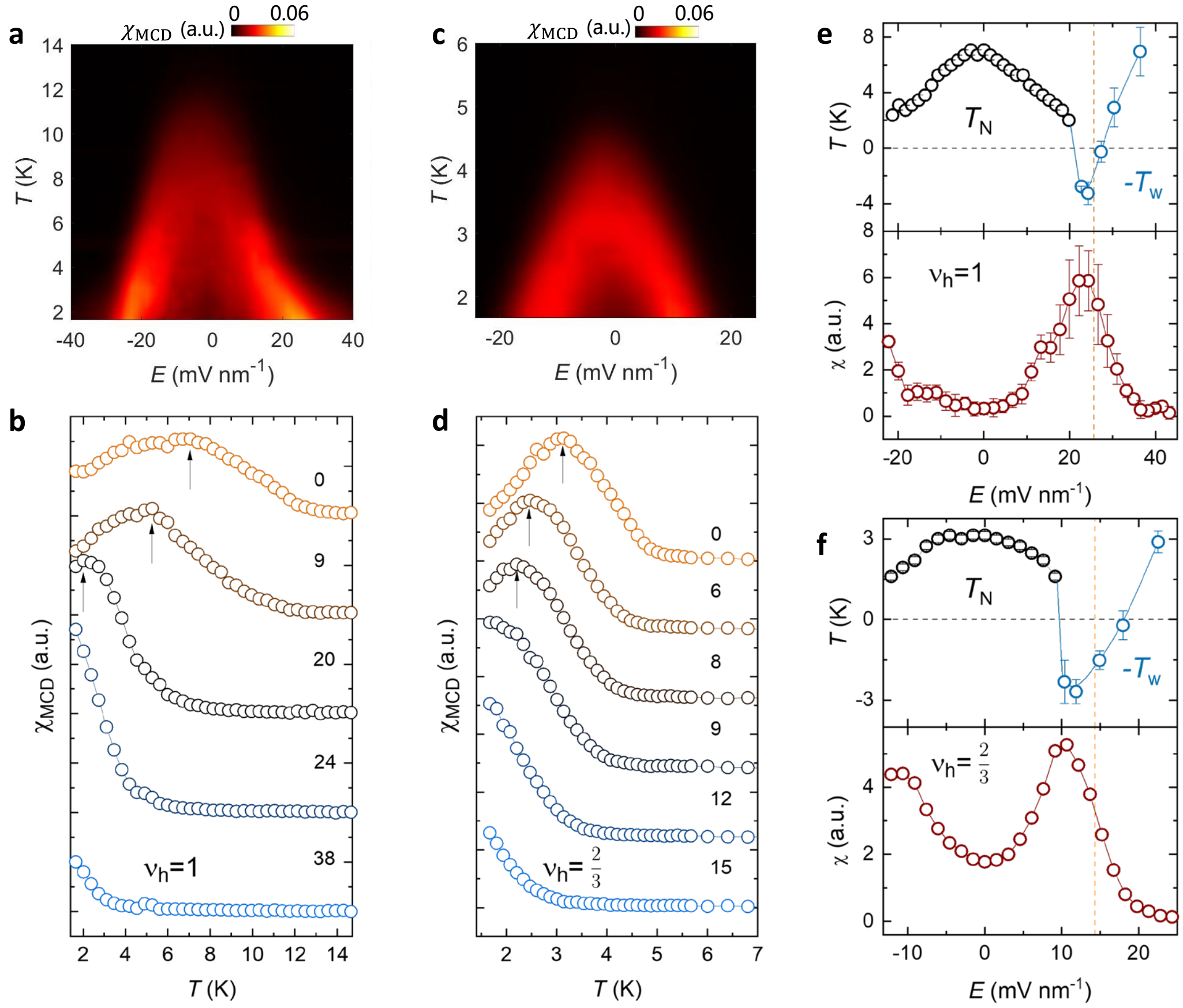

**Fig. 4 | Magnetic phase diagram of $\nu_h = 1$ and $2/3$.**

**a,c,** MCD as a function of $E$ and $T$ under a small $B$ at $\nu_h = 1$, 8 mT (**a**) and 2/3, 5 mT (**c**), respectively. **b,d,** MCD versus $T$ at representative $E$ at $\nu_h = 1$ (**b**) and 2/3 (**d**). Non-monotonic $\chi_{MCD} - T$ behaviors with pronounced $\chi_{MCD}$ peaks are observed at $E$ fields below $E_c$. The peak temperatures are assigned to be Neel temperature $T_N$, and are denoted by arrows. **e,f,** $E$ dependences of extracted $T_N$ and $-T_w$ (upper panel) and MCD susceptibility at 1.6 K (lower panel) of $\nu_h = 1$ (**e**) and 2/3 (**f**) of D(3.15°). $E_c$ obtained from the optical reflectance contrast measurements are denoted by the vertical dashed lines.

## Extended Data Figures

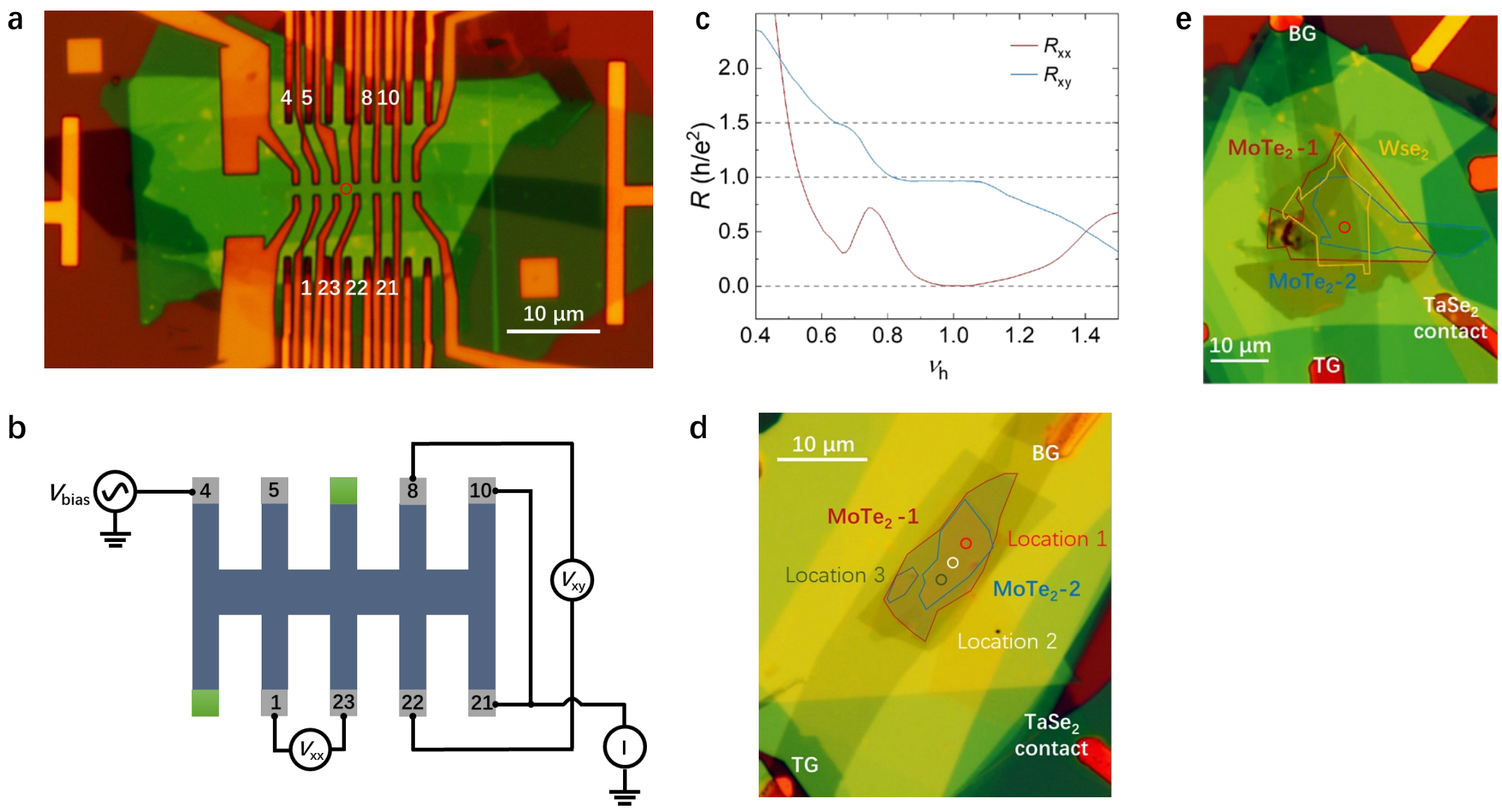

**Extended Data Figure 1 | Samples and basic characterization. a,** Optical micrograph of D(3.15°) used in this study. The scale bar is 10 μm. The Hall bar geometry is defined by standard EBL and RIE processes. Optical measurement data are taken at the red circle position. **b,** Schematic figure of the measurement configuration. For most results presented in the main text, electrodes 10 and 21 are grounded, and electrodes 4 are used as a source electrode. The longitudinal voltage drop is measured between 1 and 23, and the transverse voltage drop is measured between 8 and 22. **c,** $R_{xy}$ and $R_{xx}$ as a function of $\nu_h$ at $T = 1.5$ K with $E = 0$ mV nm$^{-1}$. The $R_{xy}$ ($R_{xx}$) is antisymmetrized (symmetrized) at $B = 0.5$ T. $R_{xy}$ shows quantized plateaus while $R_{xx}$ shows clear dips around at $\nu_h = 1$ and $2/3$, due to the chiral edge transport. **d,** Optical micrograph of D(3.0°) used in this study. The red and blue lines outline the boundaries of the twisted bilayer $MoTe_2$ sample, respectively. Optical measurement data are taken in the homogenous region on the circled locations. **e,** Optical micrograph of D(3.1°) used in dielectric sensing study. The red and blue lines outline the boundaries of the twisted bilayer $MoTe_2$ sample, respectively. The yellow line outlines the boundaries of the $WSe_2$ sensor layer. Dielectric sensing measurement data are taken in the homogenous region at the red circle position.

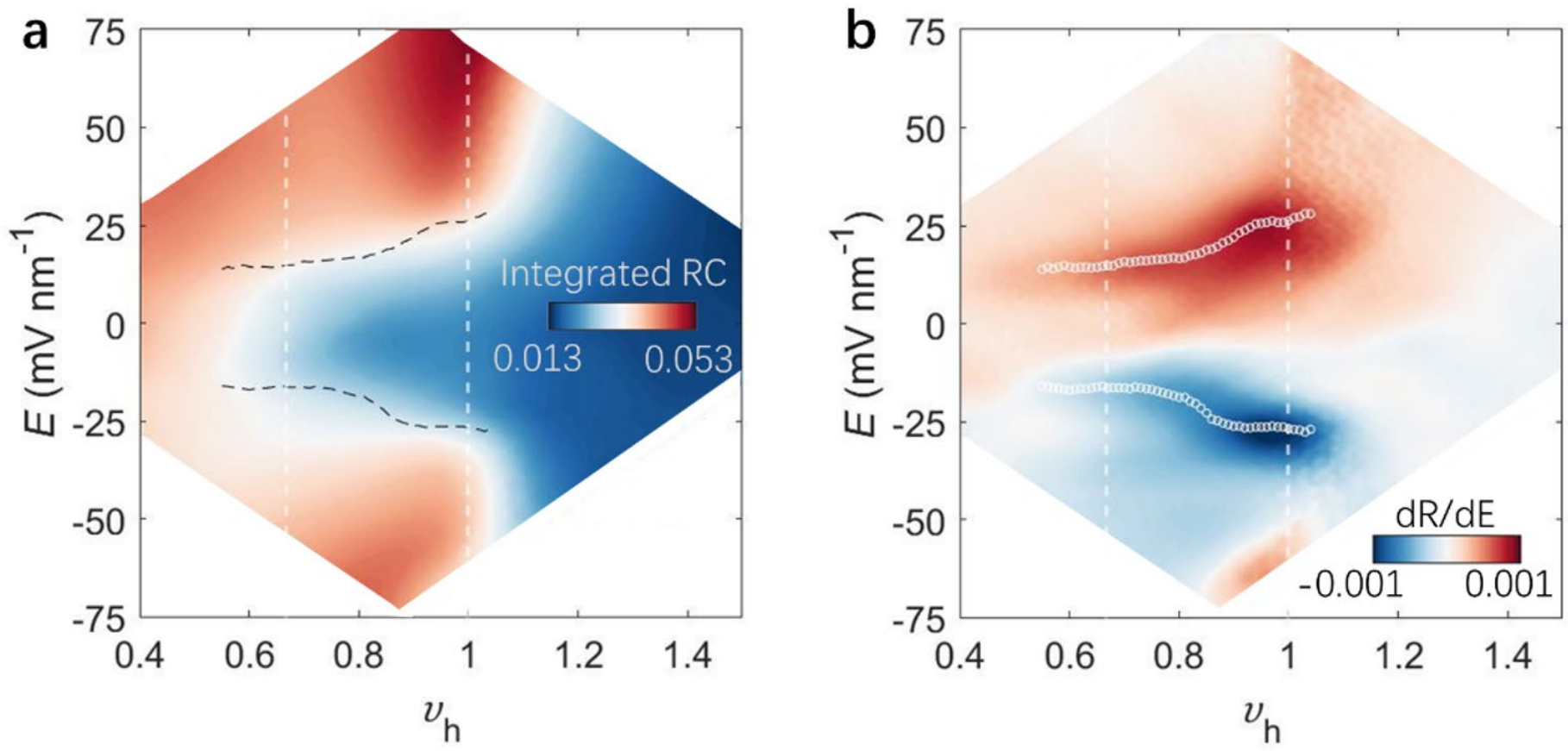

**Extended Data Figure 2 | Determination of $E_c$ of superlattice symmetry transition.**

**a,** Integrated optical reflectance contrast map as a function of $\nu_h$ and $E$ of D(3.15°). The integrated range is indicated by the *two* vertical black dashed lines in Fig. 1c. Strong integrated RC signifies that at least one of the $MoTe_2$ layers is charge neutral. Two distinct regions, layer-hybridized and layer-polarized regions, can be identified. $E_c$ of moiré superlattice symmetry transition, as determined in (**b**), is indicated by the black dash lines. **b,** The electric field derivative of the integrated RC plotted as a function of $\nu_h$ and $E$. The white circles mark the maximums and minimums at varies $\nu_h$, which are defined as the phase boundary between the layer-hybridized and layer-polarized regions, i.e., $E_c$ of lattice symmetry transition, as shown in the magnetic phase diagram in Fig. 1e.

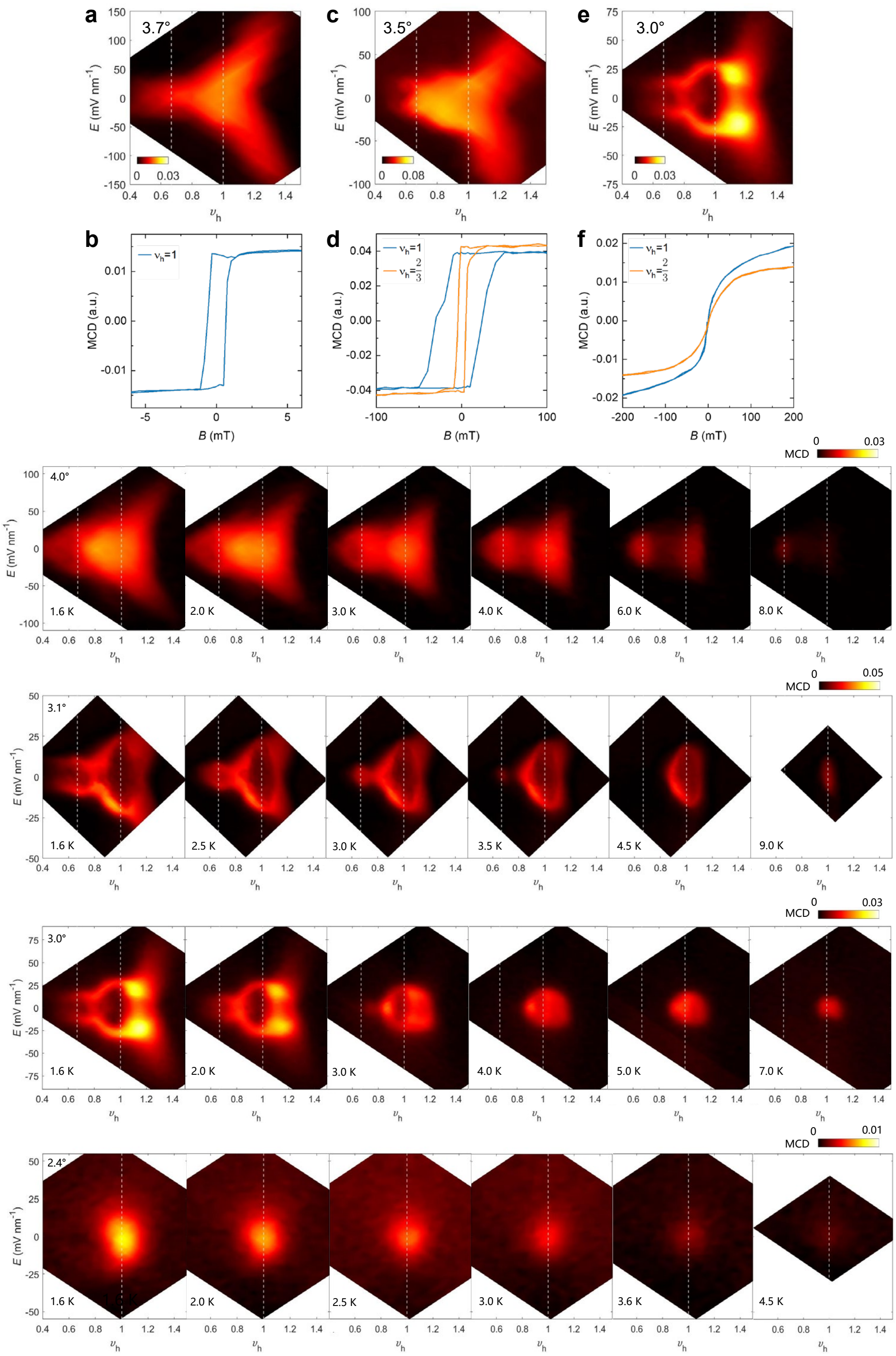

**Extended Data Figure 3| MCD characterizations of samples with different twist angles. a,c,e,** MCD maps as a function of $\nu_h$ and $E$ of D(3.7°) (**a**), D(3.5°) (**c**), and D(3.0°) (**e**), respectively, under a small $B$. $\nu_h = 1$ and $2/3$ are indicated by the vertical dashed lines. Measurements were done at 1.6 K. **b,d,f,** The corresponding $B$ dependence of the MCD at $\nu_h = 1$ and $2/3$. **g-j,** Evolution of MCD versus $\nu_h$ and $E$ as a function of $T$ in D(4.0°) (**g**), D(3.1°) (**h**), D(3.0°) (**i**) and D(2.4°) (**j**). All data were acquired under a small $B$ (9 mT, 9 mT, 5 mT, 8 mT). $\nu_h = 1$ and $2/3$ are indicated by the white dashed lines.

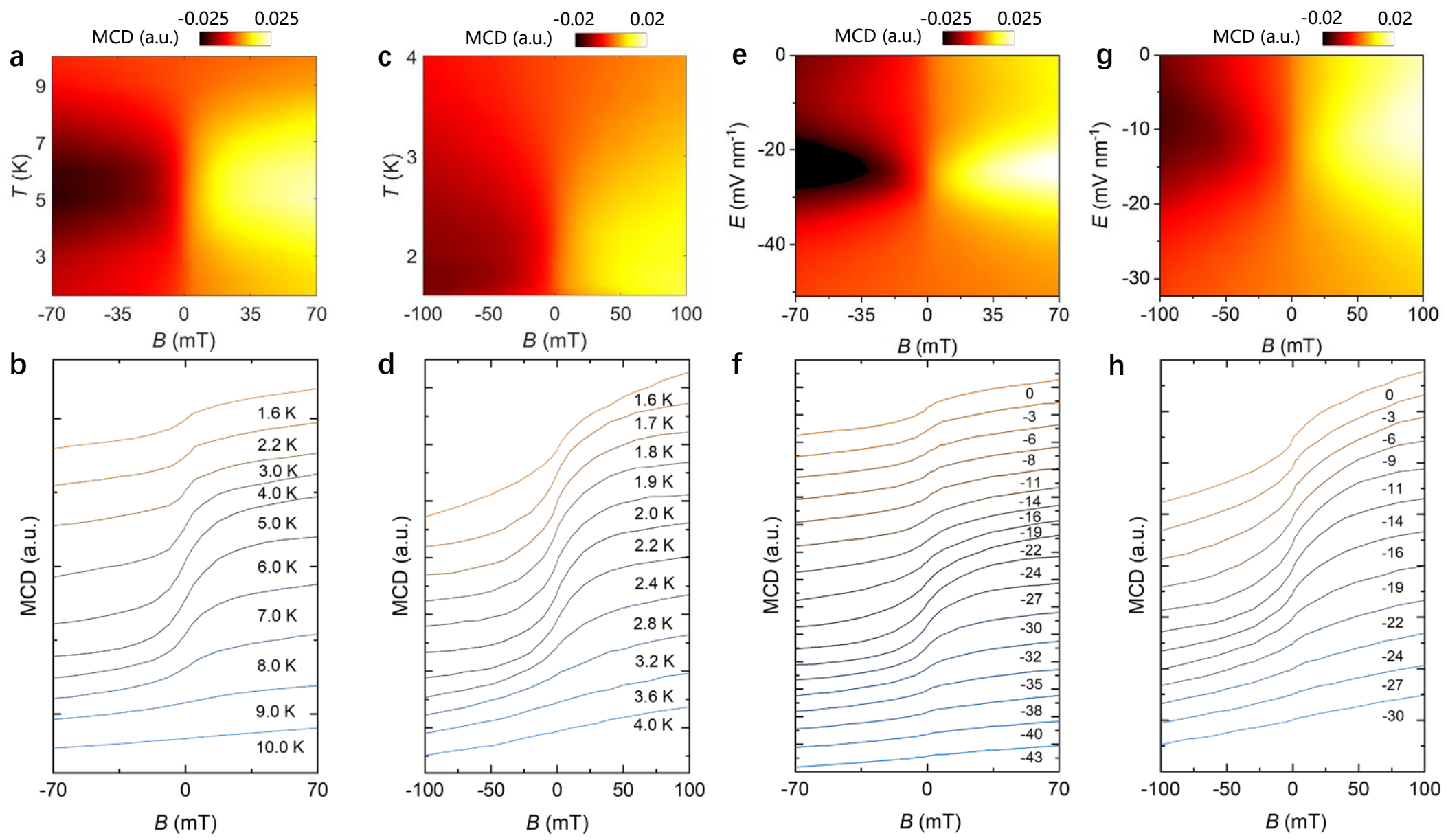

**Extended Data Figure 4 | Magnetic response of D(3.0°).**

**a,c,** MCD as a function of $B$ and $T$ at $\nu_{\mathrm{h}} = 1$ (**a**) and $2/3$ (**c**). **b,d,** MCD versus $B$ at representative $T$ at $\nu_{\mathrm{h}} = 1$ (**b**) and $2/3$ (**d**) at $E = 0$, respectively. The zero-field slops of both fillings first increase and then decrease with increasing $T$, showing a non-monotonic T dependence of $\chi$. **e,g,** MCD as a function of $B$ and $E$ at $\nu_{\mathrm{h}} = 1$ (**e**) and $2/3$ (**g**). **f,h,** $B$ dependence of MCD at representative electric fields (unit in mV nm$^{-1}$) at $\nu_{\mathrm{h}} = 1$ (**f**) and $2/3$ (**h**) of D(3.0°) measured at 1.6 K, respectively.

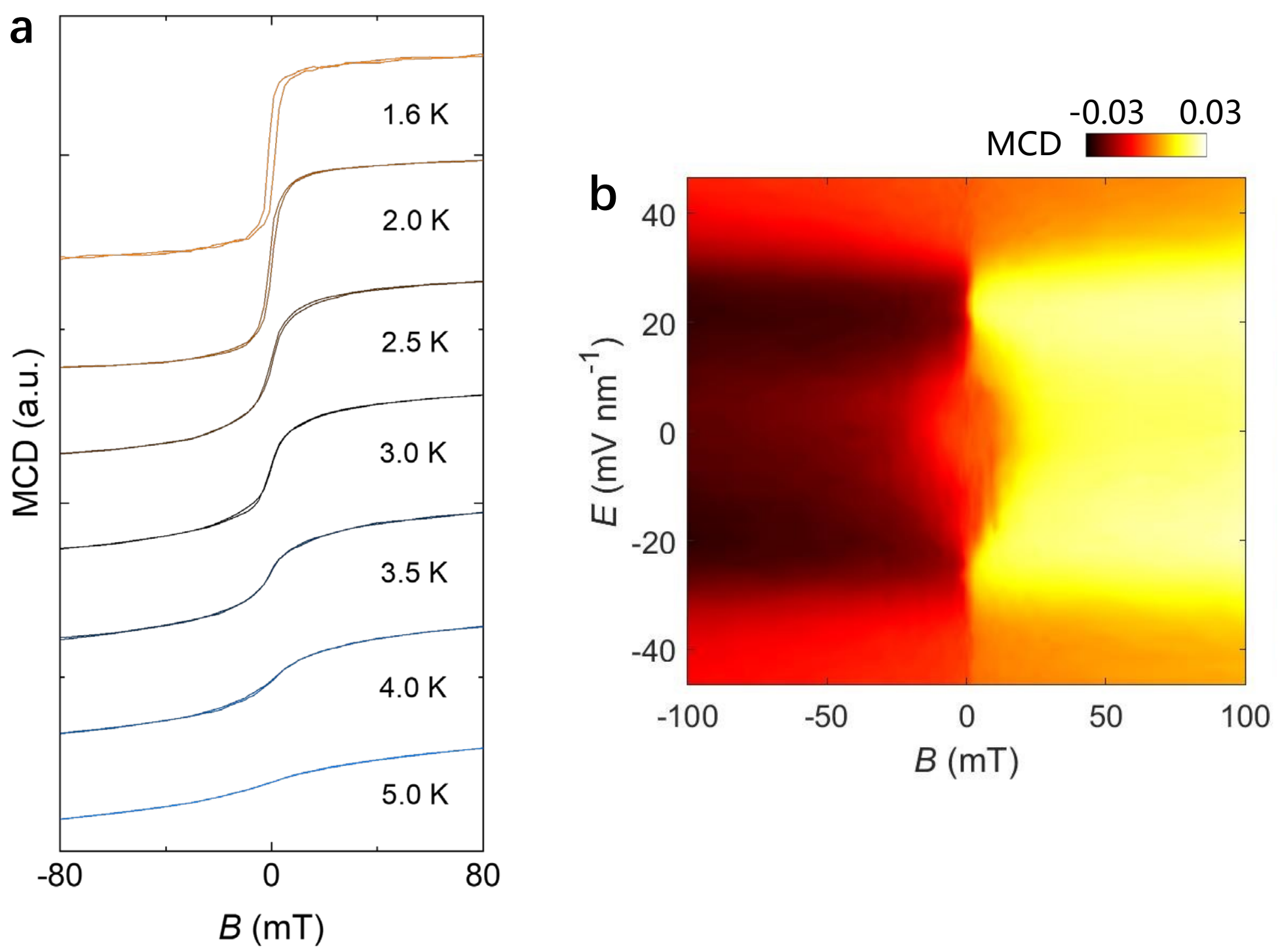

**Extended Data Figure 5 | Magnetic response of D(3.15°).**

**a,** MCD versus $B$ at representative $T$ at $\nu_{\mathrm{h}} = 1$ and $E = 24$ mV nm$^{-1}$. The zero-field slope decreases monotonically with increasing $T$. **b,** MCD as a function of $B$ and $E$ at $\nu_{\mathrm{h}} = 1$ of D(3.15°) measured at 1.6 K.

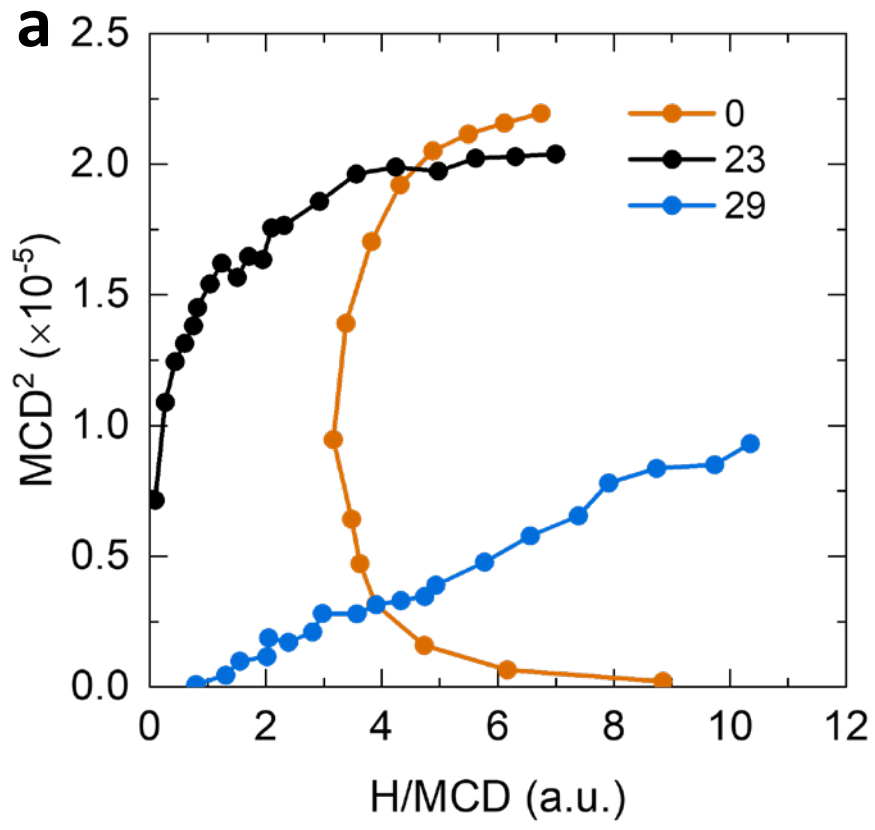

**Extended Data Figure 6 | Arrott plots of D(3.15°).** Arrott plots at representative electric fields (unit in mV nm$^{-1}$) at $\nu_h = 1$ at 1.6 K, using the data in Fig. 3a. MCD was used in place of magnetization.

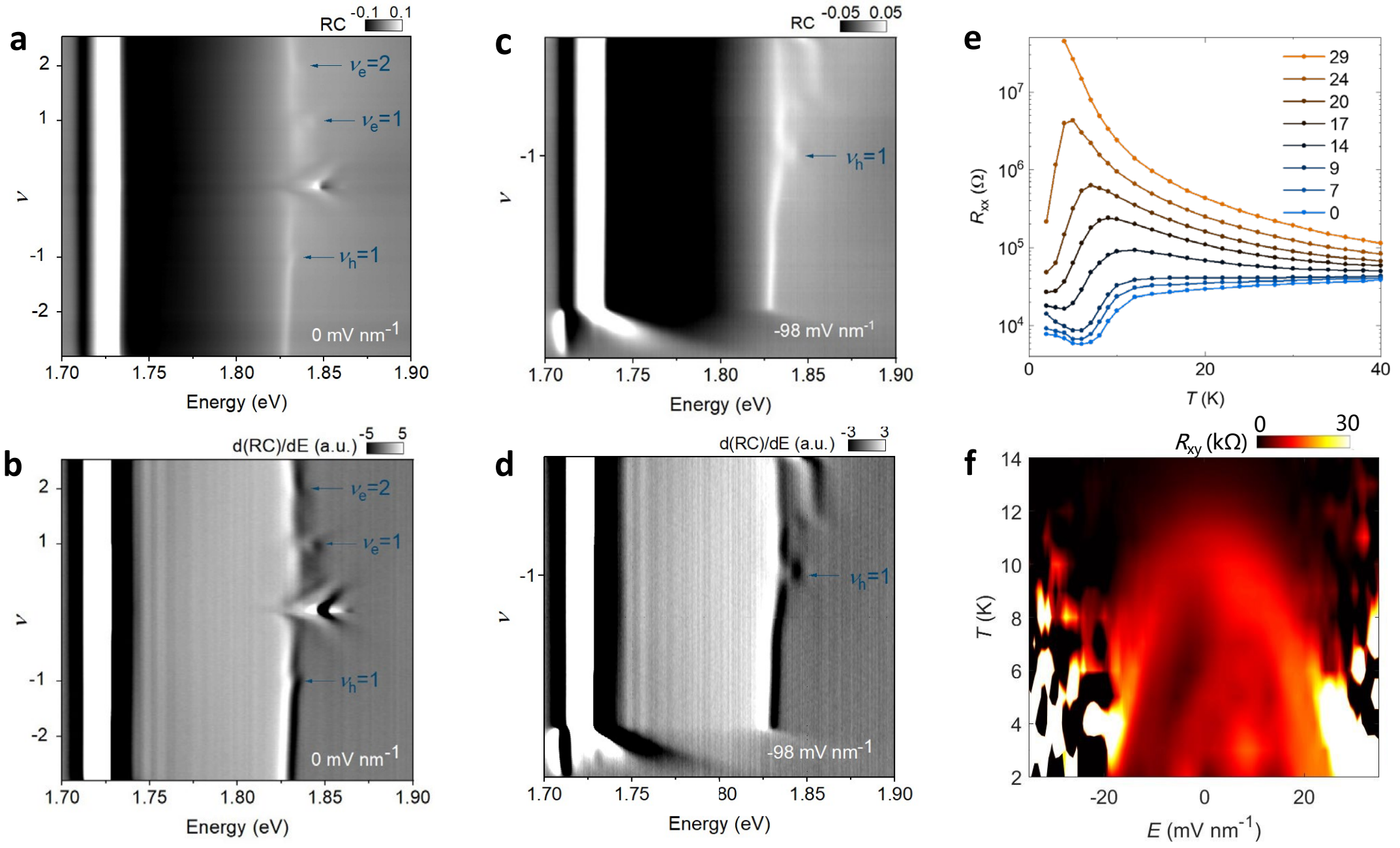

**Extended Data Figure 7 | Characterization of D(3.1°). a,c,** $\nu_h$-dependent reflection contrast spectrum showing the 2s exciton of the $WSe_2$ sensor of D(3.1°), measured at $E = 0$ and $-98$ mV nm$^{-1}$ at 1.6 K. The arrows mark the correlated insulating states identified from the blue-shift of the 2s exciton resonance. **b,d,** The corresponding differential reflection contrast spectrum of **a** and **c**, showing the pronounced feature of Mott insulating state. **e,** $R_{xx}$ vs $T$ at various $E$ of D(3.15°) under zero magnetic field. **f,** $R_{xy}$ as a function of $E$ and $T$ under a small $B$ (30 mT) of $\nu_h = 1$ at T=1.5 K**.**

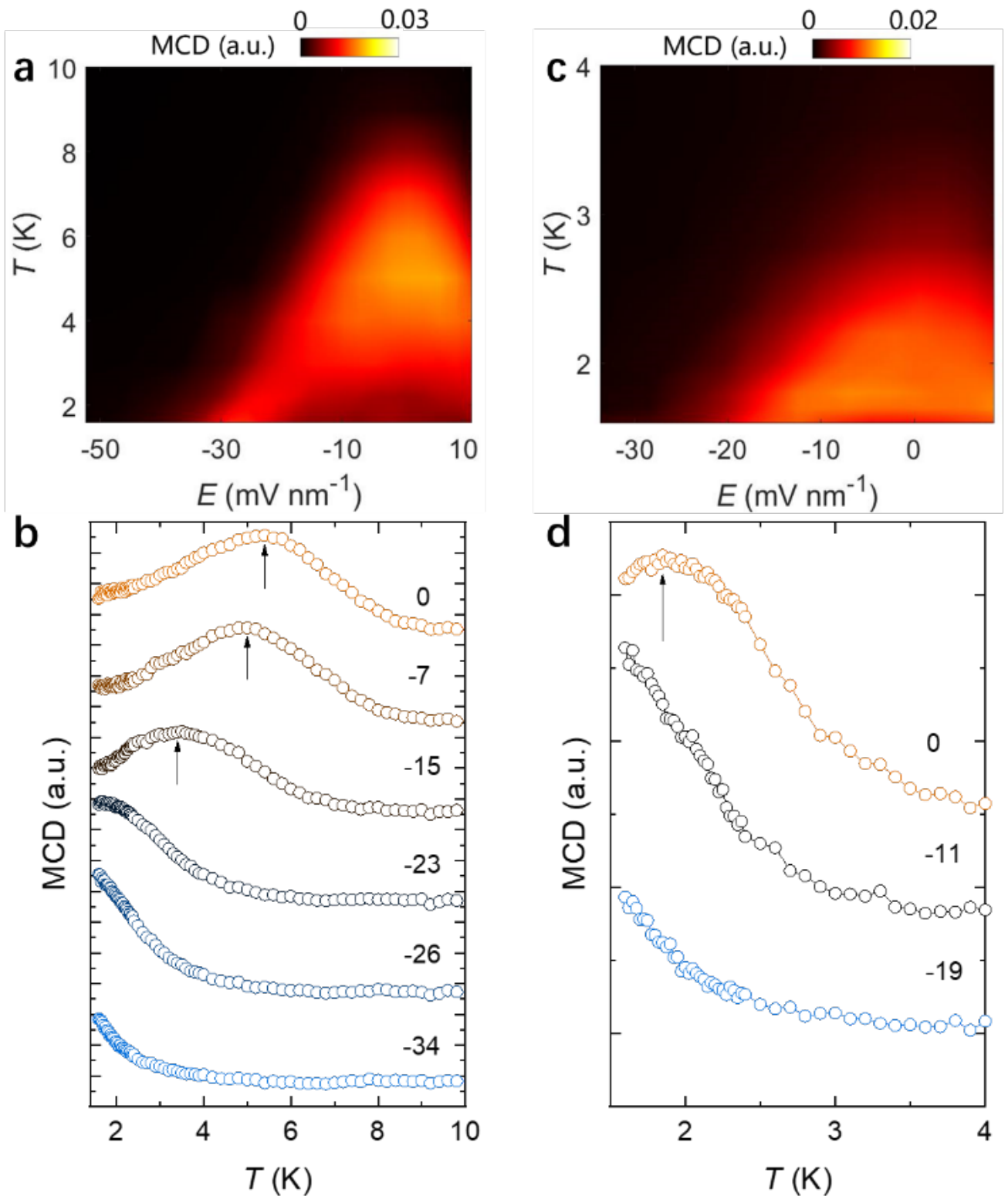

**Extended Data Figure 8 | MCD susceptibility of D(3.0°).**

**a,c,** MCD as a function of $E$ and $T$ under a small $B$ at $\nu_{\mathrm{h}} = 1$, 8 mT (**a**) and $2/3$, 8 mT (**c**) of D(3.0°). **b,d,** MCD versus $T$ at representative $E$ at $\nu_{\mathrm{h}} = 1$ (**b**) and $2/3$ (**d**). Non-monotonic $\chi - T$ behaviors with pronounced $\chi$ peaks are observed at $E$ fields below $E_{\mathrm{c}}$. The peak temperatures are assigned to be Neel temperature $T_{\mathrm{N}}$, and are denoted by arrows.

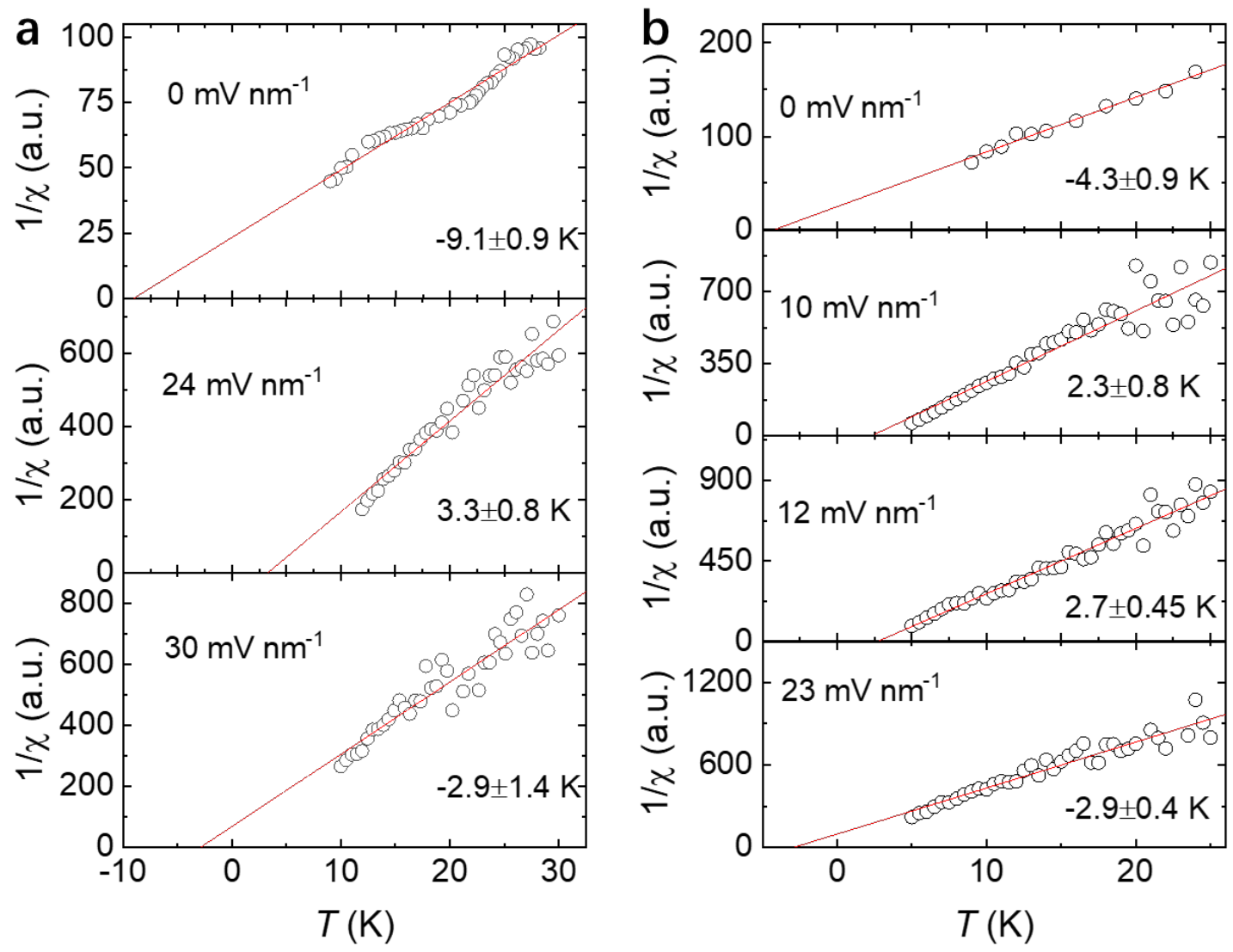

**Extended Data Figure 9 | MCD susceptibility measurement of D(3.15°).**

The temperature dependence of the inverse MCD susceptibility (symbols) and the Curie–Weiss analysis (lines) at representative electric fields at $\nu_h = 1$ (**a**) and $2/3$ (**b**), respectively. The Curie–Weiss temperature $T_w$ corresponding to the best fits are included.

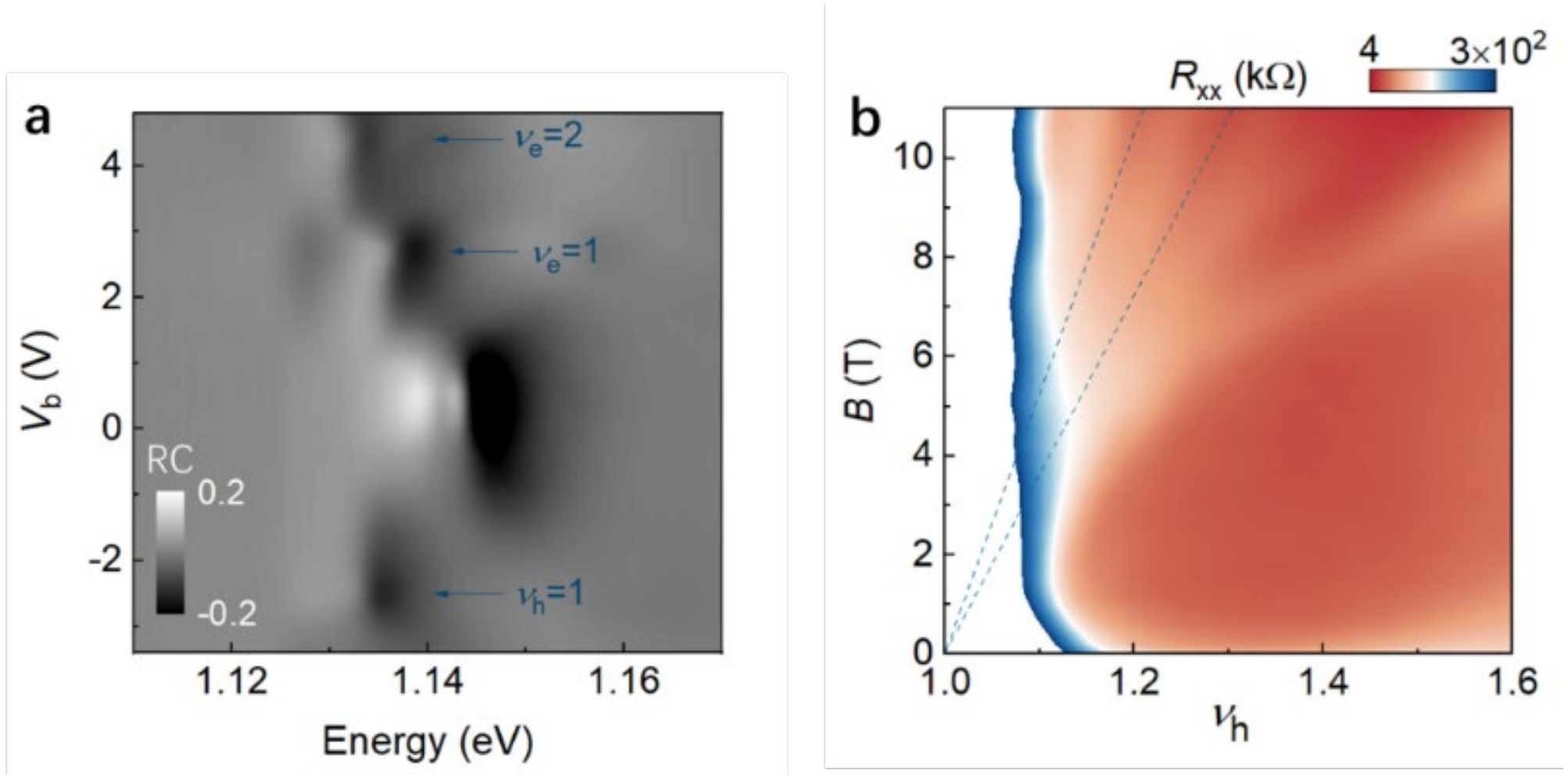

**Extended Data Figure 10 | Twist angle calibration.**

**a,** RC spectrum as a function of gate voltages at 1.6 K of D(3.0°). RC enhancement is observed at integer filling factors and denoted by the arrows. **b,** $R_{xx}$ versus out-of-plane $B$ and $\nu_h$ at $E = 57$ mV nm$^{-1}$ at 1.5 K. The measurement configuration is shown in Extended Data Fig. 1b. Dashed lines denote the Landau level $\nu_{LL} = 2$ and 3 emerging from $\nu_h = 1$ above $B \approx 7$ T.

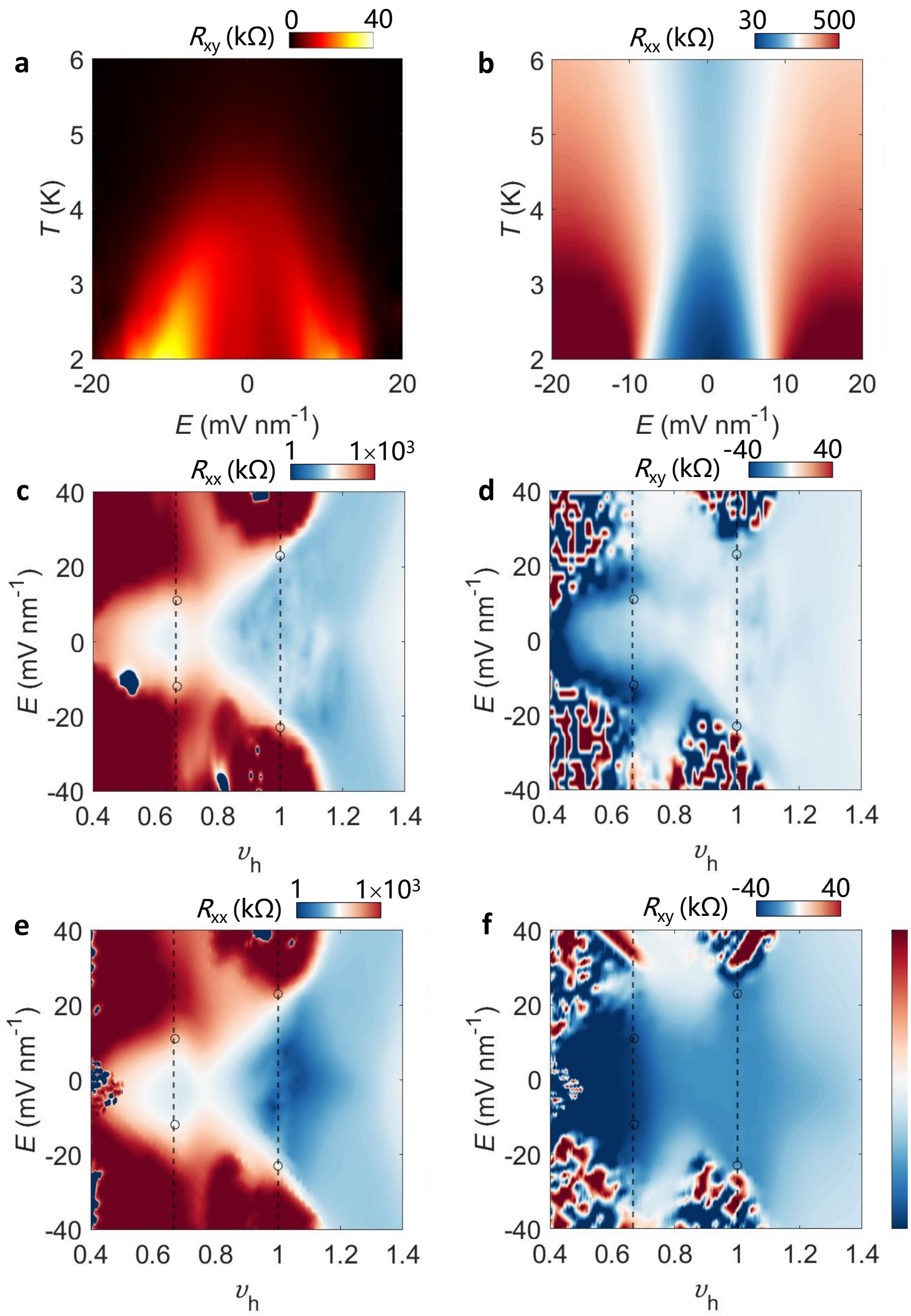

**Extended Data Figure 11 | Transport characterization of D(3.15°).**

**a,b,** $R_{xy}$ and $R_{xx}$ as a function of $E$ and $T$ under a small $B$ (20 mT) of $\upsilon_h = \frac{2}{3}$, respectively, obtained using the contact scheme shown in Extended Data Fig.1 from device D(3.15°). **c,d,** $R_{xx}$ (**c**) and $R_{xy}$ (**d**) as a function of $E$ and $\upsilon_h$ at 1.5 K, obtained using the contact scheme in Extended Data Fig.1. $R_{xx}$ and $R_{xy}$ are symmetrized and anti-symmetrized at $B = 20$ mT. **e,f,** $R_{xx}$ (**e**) and $R_{xy}$ (**f**) as a function of $E$ and $\upsilon_h$ at 1.5 K, symmetrized and anti-symmetrized at $B = 0.3$ T.

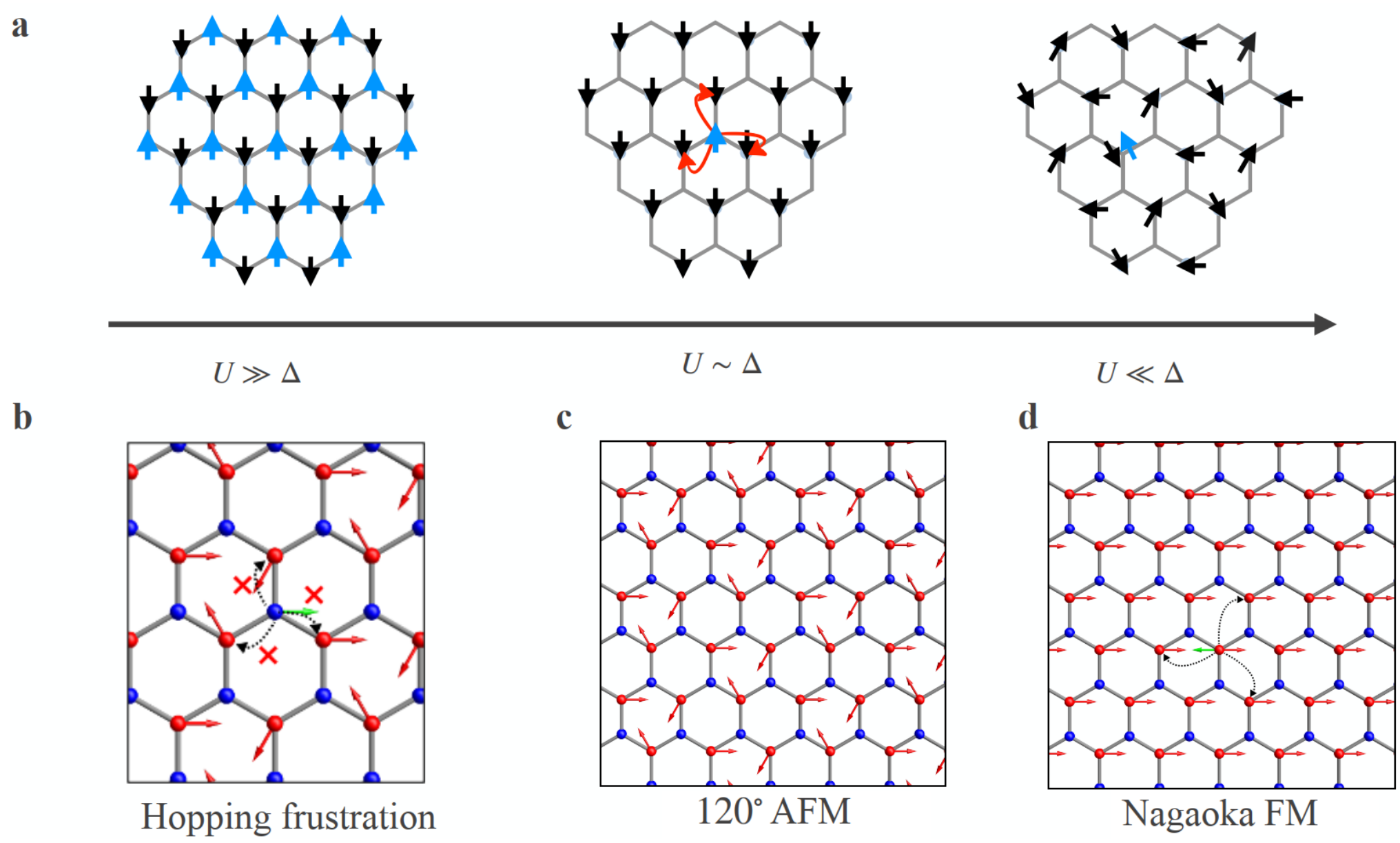

**Extended Data Figure 12 | Schematic illustration of magnetic exchange interactions. a,** The evolution of magnetic order with the displacement field is illustrated. When the electric field is strong enough, all holes occupy the A sublattice, and the effective second-nearest-neighbor (2NN) exchange drives the magnetic order to a 120° AFM configuration. As the electric field decreases near $E_c$ where the holes start to transfer into the B sublattice, the effective ferromagnetic interaction is enhanced by the direct hopping energy gain between the A and B sublattice. At a small electric field, the magnetic order is dominated by the near-neighbor exchange between the A and B sublattices, resulting in antiferromagnetism. **b,** The schematic of the Kinetic frustration of doped 120° AFM. The red crosses near the dashed arrow denote the direct hopping is suppressed due to the spin of the doped hole is not aligned with the nearest-neighbor spins. **c,** Schematic illustration of 120° AFM (left) from effective spin exchange $J \sim \frac{t_2^2}{U}$. **d,** Nagaoka FM (right) at infinite $U$ limit from direct hopping energy gain, respectively.